\documentclass[fleqn,10pt]{wlscirep}
\usepackage[utf8]{inputenc}
\usepackage[T1]{fontenc}
\usepackage{caption}
\usepackage{subcaption}
\usepackage[export]{adjustbox}

\title{The Coexistence of Infection Spread Patterns in the Global Dynamics of COVID-19 Dissemination}

\author[1,2*]{Hiroyasu Inoue}
\author[3]{Wataru Souma}
\author[1]{Yoshi Fujiwara}
\affil[1]{University of Hyogo, Graduate School of Information Science, Kobe, 6500047, Japan}
\affil[2]{RIKEN, Center for Computational Science, Kobe, 6500047, Japan}
\affil[3]{Rissho University, Faculty of Date Science, Tokyo 1418602, Japan}

\affil[*]{inoue@gsis.u-hyogo.ac.jp}

\begin{abstract}
The novel coronavirus SARS-CoV-2, commonly referred to as COVID-19, triggered the global pandemic.
Although the nature of the international spread of infection is an important issue,
extracting diffusion networks from observations is challenging because of its inherent complexity.
In this paper, we investigate 
the process of infection worldwide, including time delays, based on global infection case data collected from January 3, 2020 to December 31, 2022.
We approach the data with a complex Hilbert principal component analysis, which can consider not only the concurrent relationships between elements, but also the leading and lagging relationships.
Then, we examine the interactions among countries by considering six factors: geography, population, GDP, stringency of countermeasures, vaccination rates, and 
government type.
The results show two primary trends occurring in 2020 and in 2021-2022
and they interchange with each other.
Specifically, European, highly populated, and democratic countries, i.e., countries with high mobility rates, show leading trends in 2020.
In contrast, 
African and nondemocratic countries show leading trends in 2021-2022, followed by countries with high vaccination rates and advanced countermeasures.
The results reveal that, although factors that increase infection risk lead to certain trends at the beginning of the pandemic,
these trends dynamically changes over time due to socioeconomic factors, especially the introduction of countermeasures.
The findings suggest that international efforts to promote countermeasures in developing countries can contribute to pandemic containment.
\end{abstract}

\begin{document}

\flushbottom
\maketitle

\thispagestyle{empty}

\section*{Introduction}

The global COVID-19 pandemic caused by the novel coronavirus SARS-CoV-2 has posed an unprecedented threat to society worldwide \cite{chakraborty2020covid,world2020clinical}. Understanding the processes guiding the spread of infection during this pandemic is crucial for the development of effective mitigation and response strategies.
However, diffusion processes are inherently complex interactions and thus pose a significant challenge for analysis.
Traditional approaches, such as regression-based methods, can
partially address this complexity \cite{wong2020spreading,bhadra2021impact,xu2020possible}.
Existing studies tried to determine the causes of infection through regression analysis.
However, the diffusion processes are the outcomes of complex, interconnected systems \cite{demertzis2020modeling,tsiotas2022understanding,cornes2022covid},
not sets of independent systems, as assumed with regression-based methods, or other commonly used methods.
Specifically,
in such complex systems, interactions change the focal actors, which is usually considered a source of endogeneity in regression analyses and is not compatible with approaches assuming independence of the variables.
Therefore, diffusion processes cannot be understood without accounting for such interactions.

The spread of COVID-19 has obviously been triggered by international travel.
However, even if the same number of people travel between two countries,
diffusion may not have the same magnitude.
The number of travellers, border control policies, quarantine policies, and other countermeasures all affect the spread of disease
\cite{chinazzi2020effect,wells2020impact,nussbaumer2020quarantine}.
In addition,
the government and culture can influence the effectiveness of implemented countermeasures \cite{cepaluni2022political,frey2020democracy,annaka2021political}.
It is extremely difficult
to clarify which of these interactions affect the spread of disease between countries and within a country.
In other words, there are vast numbers of possible factors governing diffusion processes due to the combinations of those interactions, indicating the existence of a world so highly heterogeneous that the application of conventional techniques, such as regression-based models, becomes impractical.
On the other hand,
statistical physics-based \cite{castellano2009statistical,perc2017statistical} and network science-based methods \cite{Watts98,Centola10,Newman10,Barabasi16,Watts02}
have been applied in social science studies, especially in recent years.
The development of such methods underscores the importance of addressing intricate systems
without reducing their complexity.

Moreover, 
mathematical epidemic models, such as the susceptible--infectious--recovered (SIR) model \cite{ross1916application,ross1917application,ross1917application2,kermack1927contribution}, constitute an important strand of the literature for investigating the diffusion of infectious diseases.
Recent research on mathematical epidemic models has emphasized the importance
of incorporating interactions and networks in such models \cite{bansal2007individual,newman2002spread,castellano2010thresholds,fanelli2020analysis}.
Such models have greatly advanced COVID-19-related research
\cite{vespignani2020modelling,liang2020mathematical}.
However, mathematical models have limitations, including
arbitrariness and validation difficulties.
Furthermore,
while data-driven approaches, such as statistical models, have been applied to diffusion processes,
elucidating the complex behaviour of the data is still challenging.
However, a few approaches have been developed to address this issue.

In addition,
in statistical models,
the diffusion of infections with 
time series data 
is commonly investigated with 
spatiotemporal pattern analyses
\cite{castro2021spatiotemporal,kim2020spatiotemporal,alkhamis2020spatiotemporal}.
However, these studies only describe the data or apply clustering algorithms to separate areas.
They do not consider the essential part of the interaction, i.e., the delay involved in the diffusion process.

In statistical models,
principal component analysis (PCA) \cite{pearson1901liii,hotelling1933analysis} is a widely adopted technique for reducing dimensionality and recognizing patterns in various fields, including epidemiology.
When PCA is applied to global diffusion data,
patterns of disease spread, i.e., how infection cases in different countries change synchronously;
can be elucidated.
However, diffusion processes inherently involve delays in interactions.
For example, some countries may be considered spreaders while
others are considered receivers in a diffusion process.
The number of infections in the latter countries obviously increases after the number of infections rises in the former countries.
This is not a synchronous process but
a process involving time delays.
Therefore, conventional PCA is not suitable for interactions with delays.
To address this limitation, complex Hilbert principal component analysis (CHPCA)
\cite{horel1984complex} was developed as a promising approach to analyse data with a time delay.
CHPCA can capture patterns with both amplitude and phase information in time series data by introducing the Hilbert transform and using analytical signals.
The CHPCA has been widely applied in climate research \cite{rasmusson1981biennial,barnett1983interaction,hannachi2007empirical,stein2011phase} and economic research \cite{arai2013complex,Souma2022complex,vodenska2016interdependencies}.
However, CHPCA has not yet been applied to data recording the spread of infection.
Moreover, this approach could be useful for examining complex interactions in time series infection data.
In summary,
the infection spread network involving lead and lag times has not been comprehensively extracted from the observations by commonly used methods, but CHPCA is a potent method to address this concern.

One may think that simple cross-correlation analyses, such as
Granger causality analysis\cite{granger1969investigating}, can be applied to evaluate the lead and lag relationships between different sections and identify the maximal correlation coefficients for the time-shift combinations.
Presumably, with this information, we could identify significant correlations between two sections with temporal shifts.
However, there are 230 country combinations in this study, and such groups of correlations are not easily interpretable, especially since these network-like relationships may contradict each other.
In contrast, the CHPCA approach enables the lead and lag relationships to be analysed without considering pairwise combinations.
To comprehensively identify these co-movements, the CHPCA approach clearly has advantages over other methods.

Based on the above background,
we applied CHPCA to global COVID-19 infection data to investigate 
international interaction patterns in the spread of infection,
including not only the magnitude but also the lead and lag times.
Then, we examine the interaction patterns 
from the viewpoints of regions, population, restrictions and countermeasures, vaccination rates, and government.

As an overall result,
we find that 
two primary trends occurred in 2020 and in 2021--2022
that interchange between each other and 
can be characterized by the above viewpoints.
Such comprehensive trend analyses can be performed only by CHPCA.
Understanding the trends is useful for 
evidence-based policy-making in fields, such as
the containment and management of infections, especially when considering the situations of other countries.

\section*{Data}

We utilize data obtained from
the WHO Coronavirus (COVID-19) Dashboard \cite{WHOCOVID23}.
The data were collected from January 3, 2020, to January 5, 2023, 
i.e., 1,033 days.
The total number of cases over this period was 657,155,169.
The data are reported by country and day, and there are 253,869 records.
In this dataset, 92,558 records included zero or negative numbers of cases, with negative numbers reflecting adjustments in the data up to the previous day.
Instead of removing these data, we replaced these records with 0.1
since our analysis involved calculating the log ratio.
Although the original dataset includes 237 countries,
we use 231 countries in our analysis because six countries have very small populations and do not have country codes defined by ISO 3166-1.
In the CHPCA analyses, we use the data collected from 2020 to 2022, i.e., 1,028 days, and do not consider the data collected in 2023.
The data from 2023 were used to check the consistency of the data.

\begin{figure}[ht]
\centering

    \centering
    \begin{subfigure}[t]{0.45\textwidth}
        \centering
        \includegraphics[width=.95\linewidth,valign=m]{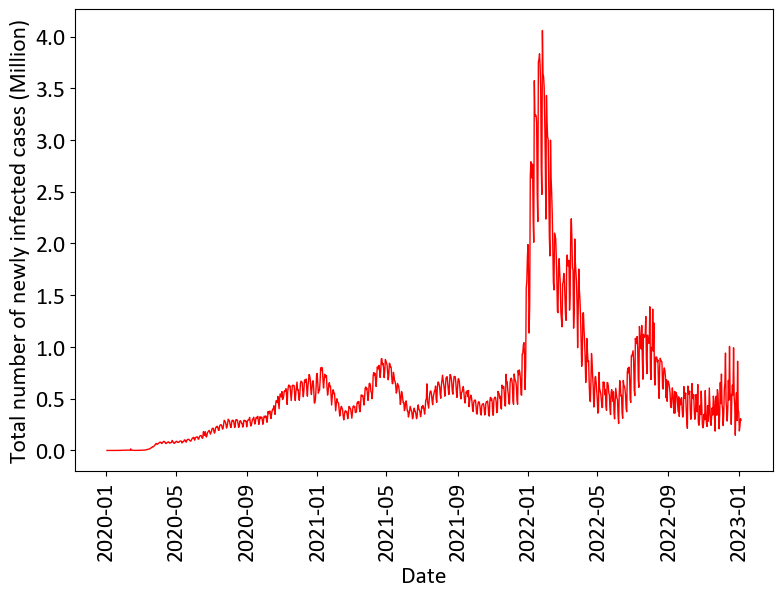}
        \caption{Total}
    \end{subfigure}
    \begin{subfigure}[t]{0.45\textwidth}
        \centering
        \includegraphics[width=.95\linewidth,valign=m]{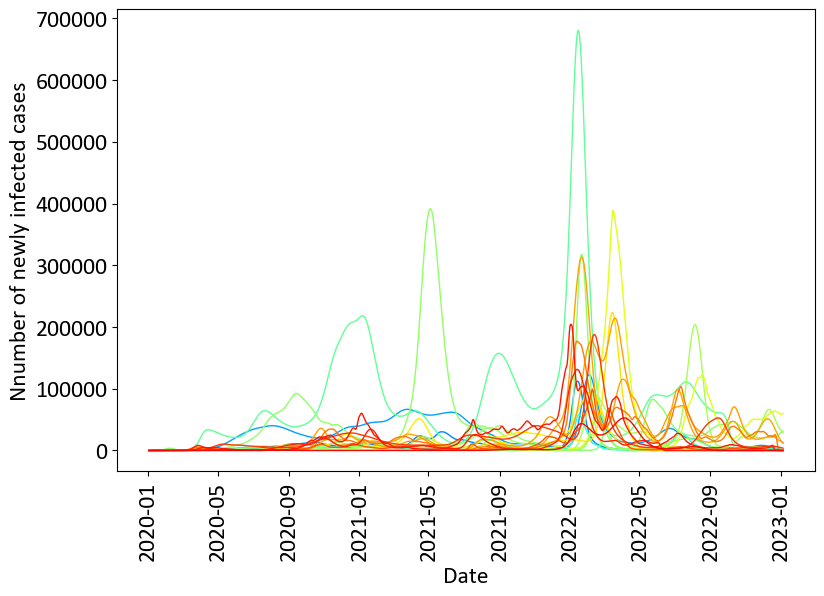}
        \caption{Cases by country (20 most populated countries)} 
    \end{subfigure}
    \hfill

    \centering
    \begin{subfigure}[t]{0.45\textwidth}
        \centering
        \includegraphics[width=.95\linewidth,valign=m]{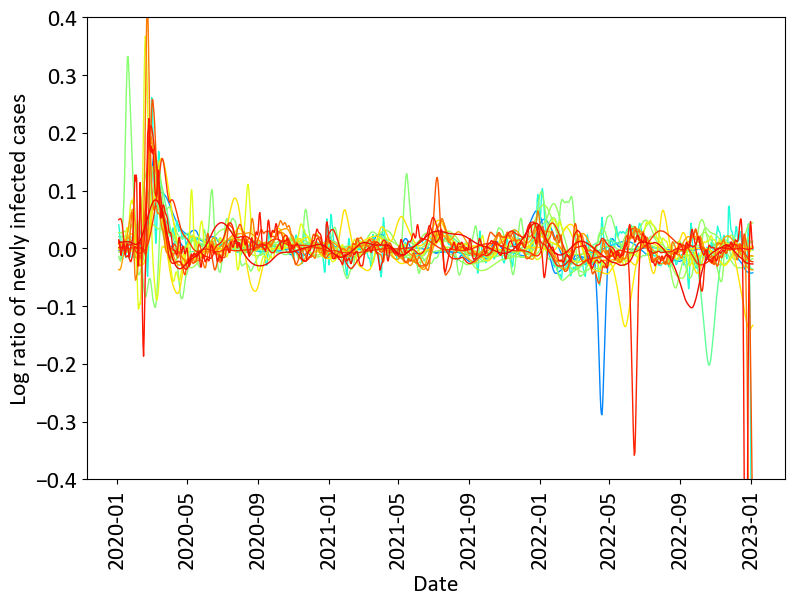}
        \caption{Log ratio (20 most populated countries)} 
    \end{subfigure}
    \begin{subfigure}[t]{0.45\textwidth}
        \centering
        \includegraphics[width=.9\linewidth,valign=m]{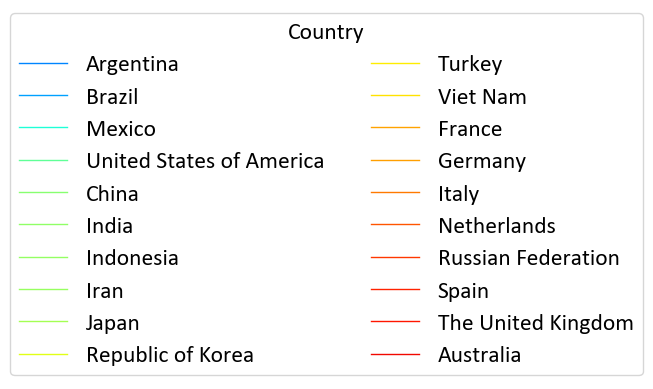}
        \caption{Legend for Panels (b) and (c)}
    \end{subfigure}
    \hfill
\caption{Number of infected cases. Panel (a) shows the daily total number of infections worldwide. Panel (b) shows the number of cases in the top 20 most populous countries after removing the weekly trends with the state-space model, and Panel (c) shows the log ratio of the results in Panel (b). Panel (d) shows the legend for Panels (b) and (c).
For Panels (b) and (c), the same figures for all countries are shown in SI Figures 1 and 2.}

\label{fig:numberofinfected}
\end{figure}

By visualizing the number of infections (Figure \ref{fig:numberofinfected}(a)), we can see
approximately six waves of infections, and the number of cases peaked around January 2022.
In addition, there are obvious regular weekly cycles in the data.
Therefore, we remove these cycles by using a state-space model \cite{kitagawa1981nonstationary} and visualize
the daily number of cases for each country after removing the weekly trends
Importantly, the data show that
the waves in different countries are not synchronized (Figure \ref{fig:numberofinfected}(b)
for the 20 most populous countries and SI Figure 1 for all 231 countries); thus, PCA is not suitable for analysing these data, as discussed in the Introduction.

To evaluate the temporal differences in wave development among countries,
we take the log ratio for each country, which is
defined by the following equation (Figure \ref{fig:numberofinfected}(c) for the 20 most populous countries and SI Figure 2 for all 231 countries):
\begin{equation}
r_{c,t}=\log(x_{c,t})-\log(x_{c,t-1}),\label{eqn:logratio}
\end{equation}
where $c$ represents a country ($c=1, 2, ... C$), $t$ is a sequential date ($t=1, 2, ... T$), and $x$ is the number of infected cases.
In this paper, $C = 231$ and $T=1,028$ for the analysis of the entire period,
and $T=365$ for the yearly analyses.
The wave development in the different countries is not synchronized, as shown in Figure \ref{fig:numberofinfected}(c).
However, wave development in different countries may show patterns in lead and lag times.
Understanding such patterns would be helpful for understanding the spread of infection,
which is the central issue of this paper.

We also use auxiliary data to interpret the causes of 
co-movements in the infection data.
These data include population and GDP information \cite{worldbankpop22}, the stringency index and containment and health index \cite{ourworldindata23}, the vaccination rate \cite{ourworldindata23}, and the democracy index \cite{economistintelligence22}.
All of these data have enough observations (SI Table 1) to be combined with the data of the 231 countries used in our analyses.
However, the following records are not available:
twenty countries lack population data,
forty-eight countries lack GDP data,
fifty countries lack stringency index and 
containment and health index data,
seventeen countries lack vaccination data,
and sixty-six countries lack democracy index data.
Although of course it is understandably better to have complete datasets,
it can be generally said that those countries that are excluded do not have critical importance to the total dataset. Therefore, we consider the missing data in these sets to be acceptable.

The stringency index was proposed by Our World in Data \cite{ourworldindata23},
and it is the composite of the following nine metrics: school closure, workplace closure, cancellation of public events, restrictions on public gatherings, public transport closures, stay-at-home requirements, public information campaigns, restrictions on internal movement, and international travel restrictions \cite{Hale2020}.
The containment and health index, which was also proposed 
by Our World in Data \cite{ourworldindata23}, is based on the stringency index and considers the testing policies, the extent of contact tracing, the requirements to wear face coverings, and vaccination policies.
The difference between the stringency index and the containment and health index concerns the application of advanced prevention measures.

The democracy index was proposed by
the Economist Intelligence Unit \cite{economistintelligence22}.
It is a composite index based on five categories: electoral process and pluralism, government functions, political participation, political culture, and civil liberties.

We examine the eigenvectors based on these auxiliary data
but use GDP per capita instead of GDP and the vaccination per capita instead of the actual vaccination rate.
Although the stringency index and the containment and health index are similar and have relatively strong correlations,
the other data are not correlated (SI Figure 4).

\section*{Methods}

We first introduce the complex correlation matrix.
Then, we describe the rotational random shuffling method,
which produces null models to assess the significance of the eigenvectors.

\subsection*{Complex correlation matrix}

We use complex Hilbert principal component analysis (CHPCA) \cite{horel1984complex} approach 
to analyse the co-movements of countries in the time series data.
This method enables us to reveal the relationships in the data not only at the same points in time but also with lead and lag times.
Unlike non-time series data, in which observations are independent and there is no temporal order, time series data can have dependencies; i.e., the observation targets can affect each other.
If we were to apply principal component analysis (PCA) to the time series data,
we would observe only the co-movements that occur at the same time.

There are different methods other than the CHPCA to examine causality on time series data, such as Granger causality~\cite{granger1969investigating}.
However, in terms of the interpretability of the results and the calculation time, these other methods have limitations, which has been explained in the Introduction.

We do not use statistical software or libraries to perform the CHPCA.
Instead, we use our own code.
All of the procedures involve mathematical operations, and there are no unclear points.

We start with the time series data of the log ratio $r_{c,t}$ shown in Eqn.(\ref{eqn:logratio}) and calculate the standardized log ratio.
The mean of the log ratio is
\begin{equation}
\langle r_{c} \rangle=\frac{1}{T}\sum_{T}^{t=1}r_{c,t}
\end{equation}
and the variance is
\begin{equation}
\sigma^{2}_{c}=\frac{1}{T}\sum^{T}_{t=1}(r_{c,t}-\langle r_{c} \rangle).
\end{equation}
Then, we obtain the standardized log ratio as follows:
\begin{equation}
w_{c,t}=\frac{r_{c,t}-\langle r_{c} \rangle}{\sigma_{c}}.\label{eqn:standard}
\end{equation}

Next, we calculate the analytical signal of the standardized log ratio, which is defined as follows:
\begin{equation}
\tilde{w}_{c,t}=w_{c,t} + i \mathcal{H}[w_{c,t}],\label{eqn:analytical}
\end{equation}
where $\mathcal{H}$ is the Hilbert transform and $i$ is the imaginary number.

A matrix that has $\tilde{w}_{c,t}$ as a component, i.e.,
\begin{equation}
\mathbf{W}=[\tilde{w}_{c,t}],
\end{equation}
can be used to obtain a complex correlation matrix, which can be expressed as follows:
\begin{equation}
\mathbf{C}=\frac{1}{T}\mathbf{W}\mathbf{W}^*,
\end{equation}
where $\mathbf{W}^*$ is an adjoint matrix, i.e., the transpose and complex conjugate of $\mathbf{W}$.
$C_{ab}$ is an element of $\mathbf{C}$, which can be written as
\begin{align}
C_{ab} & ={\rm Re}(C_{ab})+i{\rm Im}(C_{ab}),\nonumber \\
 & = |C_{ab}|e^{i\phi_{ab}},\label{eqn:cormat}
\end{align}
where ${\rm Re}$ and ${\rm Im}$ are the real and imaginary parts of the element, respectively; $|C_{ab}|$ denotes the correlation of the amplitude, and $\phi_{ab}$ denotes the correlation of the phase.

\subsection*{Rotational random shuffling}

Unlike the commonly used PCA approach,
we utilize temporal relationships in our analyses.
In the conventional null model approach,
we can randomly take a sample from $w_{c,t}$ for a country $c$ and perform PCA
to create a null model, expressed by the following:
\begin{equation}
w_{c,t} \rightarrow w_{c,{\rm rand}[1,T]},\label{eqn:random}
\end{equation}
where ${\rm rand}[1,T]$ is a random integer from 1 to T.
Note that random sampling needs to be performed without duplication for all $t$.
Then, we can create an error interval using a group of null models.
However, this process disrupts the temporal relationships in the data.
Thus, we use the rotational random shuffling (RRS) method developed by
Iyetomi et al. \cite{iyetomi2011causes,iyetomi2011fluctuation} as follows.

The RRS method can be modified from the form shown in Eq. (\ref{eqn:random}) and expressed as follows:
\begin{equation}
w_{c,t} \rightarrow w_{c,{\rm mod}[t+\tau,T]},
\end{equation}
where $\tau$ is a random number between 0 and T. With this approach, $w_{c,t}$ and $w_{c,t+1}$ are randomly changed into $w_{c,{\rm mod}[t+\tau,T]}$ and $w_{c,{\rm mod}[t+1+\tau,T]}$, respectively.
This shuffling approach reserves the temporal relationships and autocorrelation in the time series data of country $c$ and enables us to randomize cross-sectional correlations.

\subsection*{Complex Hilbert principal component analysis}

Based on the approaches presented thus far,
we can perform PCA based on the complex correlation matrix and identify significant eigenvectors.
The complex correlation matrix can be decomposed into
\begin{equation}
\mathbf{C}=\sum^{C}_{j=1}\lambda_j v_j v^*_j,
\end{equation}
where $v_j$ is the eigenvector, $v^*_j$ is the adjoint vector, and $\lambda_j$ is the eigenvalue.

The eigenvalues can be sorted in descending order for the observed data and the sample data created by the RRS method.
Then, the eigenvectors at the same rank can be compared.
If there are significant eigenvalues for the observed data, these eigenvalues will be greater than the mean of the eigenvalues for the sample data, 
and we can expect that even if some eigenvalues of the observed data are significant,
most will not be.
To concretize this process, we create 20 samples via the RRS method and establish an error interval with a boundary at 2.33 times the standard error from the mean so that we can test whether the observed data are significant at the 1\% level.

The CHPCA approach has many advantages compared with other conventional methods.
If we apply PCA to time series data,
i.e., not using the analytical signals (Eqs. (\ref{eqn:analytical}) and (\ref{eqn:standard})),
we can analyse only the simultaneous co-movements that occur.
In contrast, the CHPCA approach reveals co-movements at different moments, considering leads and lags in the data.

\section*{Results and Discussion}

\subsection*{Significant eigenvectors}

\begin{figure}[ht]
\centering
\includegraphics[width=.4\linewidth]{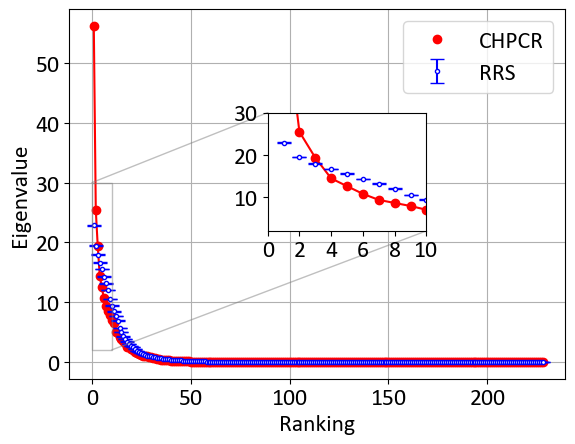}
\caption{Scree plot of CHPCA and the RRS. The data cover the entire period. The error bars for the RRS indicate 1\% significance level, i.e., 2.33 times the standard error from the mean.}
\label{fig:scree}
\end{figure}

We apply the CHPCA approach to
the infection number for the entire period from January 3, 2020, to the end of 2022 collected in the 231 countries
so that we can obtain the co-movement of the countries, including the lead and lag.
Then, we obtain three significant eigenvalues and the corresponding eigenvectors.
The ordered eigenvalues of the observations and the random samples are shown in a scree plot (Figure \ref{fig:scree}).
The eigenvalues of the observations decrease as the rank increases and are less than those of the random samples, which is understandable.
In the following sections, we discuss the eigenvectors of the first eigenvalues in detail and
the eigenvectors of the second and third eigenvalues as supplementary analyses.

We also apply the CHPCA approach separately to the data collected in 2020, 2021, and 2022.
The scree plots for the yearly data, i.e., 2020, 2021, and 2022, show different results (SI Figure 5) than the scree plot for the entire period.
Unlike in the analysis of the entire period, only two eigenvalues are statistically significant.
In addition, for the rank-one eigenvalue, the gap between the eigenvalue and the error interval is not large, especially in 2021, which suggests 
that there was no dominant trend in 2021.
In addition, for 2020, the second eigenvalue is significant but very close to the error interval, and only the first eigenvalue is significant in 2022.
Since the results suggest that 
the trends are not temporarily stable and instead change over time,
we discuss both the results of the entire period and the yearly results.

The eigenvectors in the CHPCA results show not only the magnitude, but also the temporal relationships between elements.
Therefore, the eigenvectors in the CHPCA have two dimensions.
We associate the argument with the abscissa and the amplitude with the ordinate.
On the other hand, since the original results are complex numbers, they can be plotted on complex planes (see SI Figure 7 as an example).

The eigenvector for the entire period spans approximately one year, which is one third of the entire period.
Although the eigenvector for the entire period can have indicated that a relationship existed for more than a year, there is no such a relationship,
suggesting that multiple-year-long lead and lag relationships do not exist between countries.
One remarkable point regarding the shapes of the yearly data is that the 2021 and 2022 results are spread more widely than the 2020 results.
These findings suggest that international relationships between countries had longer lead and lag times in later years.

\subsection*{Trends and transitions: interpretation by region}

Like in conventional PCA, eigenvectors require interpretation.
However, since there are 231 countries, it is not realistic to discuss each country.
Therefore, we group the countries so that we can understand the trends by groups.
First, we group the countries by region.

The first eigenvectors obtained by the CHPCA are found to be significant for the entire period and each year. 
Therefore, we have four eigenvectors of the first rank for the entire period and for each year of 2020--2022. They are grouped by region, respectively (SI Figure 6).

\begin{figure}[ht]
\centering

\begin{subfigure}[t]{0.47\textwidth}
    \centering
    \includegraphics[width=.9\linewidth,valign=m]{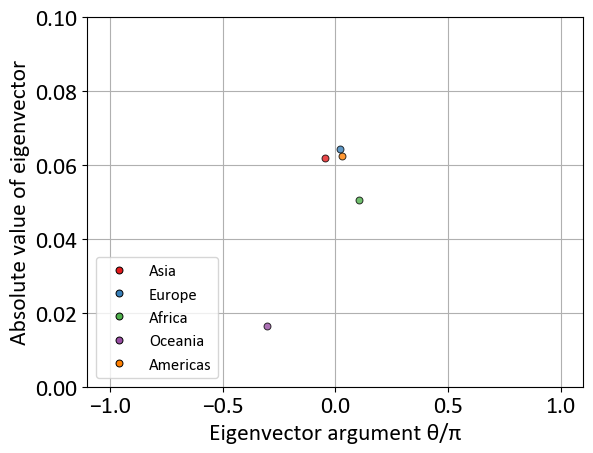}
    \caption{Entire period} 
\end{subfigure}
\begin{subfigure}[t]{0.02\textwidth}
\end{subfigure}
\begin{subfigure}[t]{0.47\textwidth}
    \centering
    \includegraphics[width=.9\linewidth,valign=m]{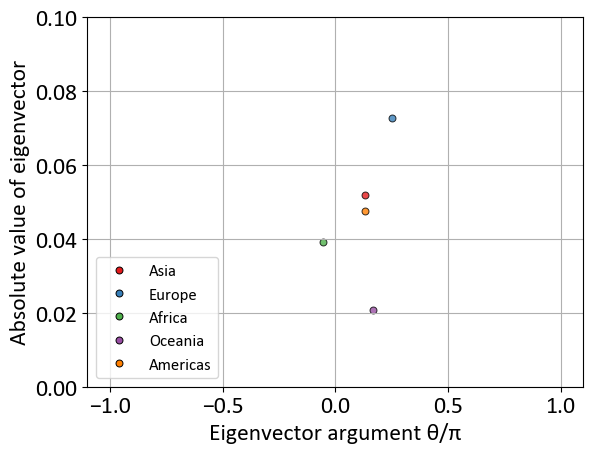}
    \caption{2020} 
\end{subfigure}
\hfill

\vspace{3ex}

\begin{subfigure}[t]{0.47\textwidth}
    \centering
    \includegraphics[width=.9\linewidth,valign=m]{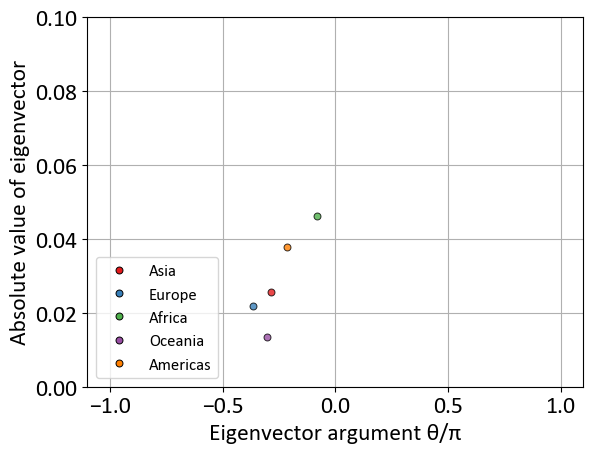}
    \caption{2021}
\end{subfigure}
\begin{subfigure}[t]{0.02\textwidth}
\end{subfigure}
\begin{subfigure}[t]{0.47\textwidth}
    \centering
    \includegraphics[width=.9\linewidth,valign=m]{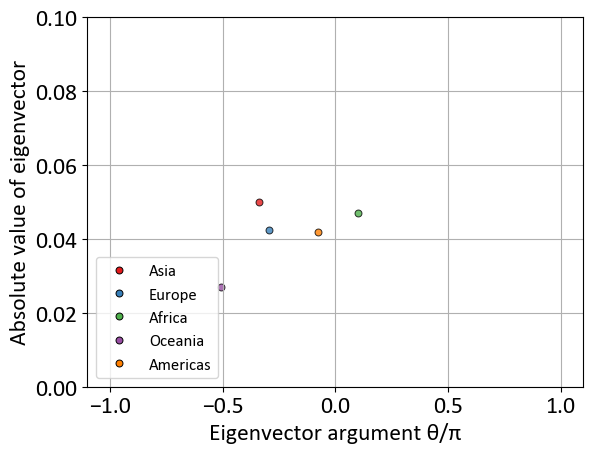}
    \caption{2022}
\end{subfigure}
\hfill

\caption{Barycentres for each region in the first eigenvectors. The abscissa corresponds to the argument, and the ordinate corresponds to the absolute value (amplitude) of the eigenvector.
Panels (a-d) represent the entire period, 2020, 2021, and 2022, respectively. Note that time progresses from right to left.}
\label{fig:eigen1regionbary}
\end{figure}

Even after grouping by region,
the plots in each region are still dispersed and it is not possible to clearly identify which regions are leading or lagging.
Therefore, to enhance interpretability,
we calculated the barycentres of the regions (Figure \ref{fig:eigen1regionbary})
and found that the lagging region is clearly Oceania but other regions, i.e., Asia, Europe, Africa, and the Americas are leading regions, although Africa is particularly leading among them.
In terms of the amplitude, the trend is the same. There is not much difference among regions, except for Oceania, which clearly has the smallest amplitude.
Importantly, this trend is not stable over time but changes as indicated by the following yearly analyses.

The trend in 2020 does not correspond to that of the entire period.
In 2020, the plots of Europe clearly have a large amplitude, while the plots of Oceania have a small amplitude. In terms of temporal relationships, Europe leads the other regions (panel (b) in Figure \ref{fig:eigen1regionbary}).

The trends in 2021 and 2022 are completely different from that in 2020.
(panels (c) and (d) in Figure \ref{fig:eigen1regionbary}). In these two years, Africa is the leading region, while Europe showed lagging relationships. No countries are prominent in terms of amplitude.

We then calculated the mean distance from the barycentre of each region
(SI Table 2).
The distance is greater in 2021 and 2022 than in 2020.
This result suggests that
the infection initially spreads regionally, which is understandable because
the countries in each region are neighbours, and people may have large amounts of traffic within each region.
However, the 2020 trend disappears over time, and simple regional proximity is not sufficient to explain the spread.

\subsection*{Trends and transitions: interpretation by population and GDP per capita}

The element of the eigenvectors can be characterized
by population and GDP.
Since the regional interpretation does not seem to be complete,
we can speculate that these trends can be interpreted
through different characterizations.

The eigenvectors can be characterized by
population (SI Figure 8).
Since population is not a categorical value, but a discrete number,
we use a continuous scale for colouring the elements.
In addition, we do not assign the actual population for each element,
but the rank of the population.
This is because the populations of the countries have highly skewed distributions.
The rank is converted to a value between 0 and 1 so that the data represent a uniform interval.
This procedure to add information by rank is also used for the following analyses presented in this paper.
Importantly, although we characterize the elements by population,
the eigenvectors are unchanged. (The eigenvector plots shown in SI Figure 6 have the exact same positions as those shown in SI Figure 8.)

When we examine the entire period,
countries with large populations have larger amplitudes than do those with smaller populations over the entire period.
This trend is also observed in 2020.
It is understandable that
larger populations are more likely to disseminate infectious diseases to other countries.
However, this trend is not observed in either 2021 or 2022.

\begin{figure}[ht]
\centering

\begin{subfigure}[t]{0.48\textwidth}
    \centering
    \includegraphics[width=\linewidth,valign=m]{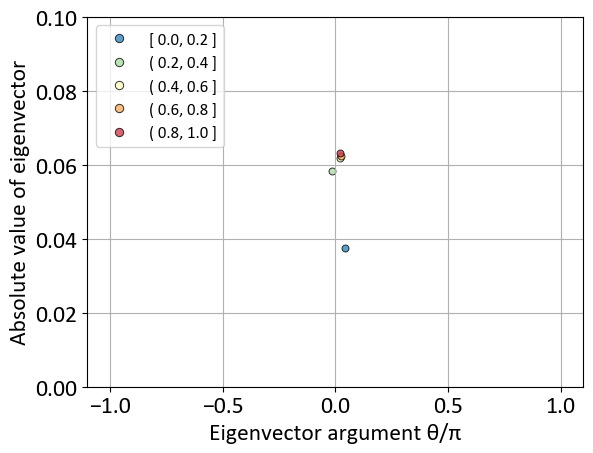}
    \caption{Entire period} 
\end{subfigure}
\begin{subfigure}[t]{0.01\textwidth}
\end{subfigure}
\begin{subfigure}[t]{0.48\textwidth}
    \centering
    \includegraphics[width=\linewidth,valign=m]{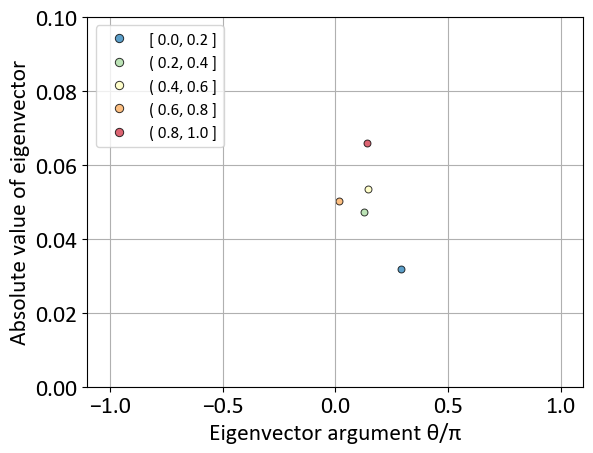}
    \caption{2020} 
\end{subfigure}
\hfill

\vspace{3ex}

\begin{subfigure}[t]{0.48\textwidth}
    \centering
    \includegraphics[width=\linewidth,valign=m]{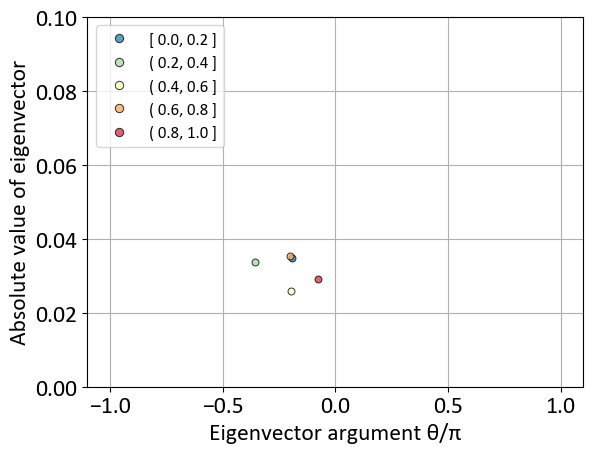}
    \caption{2021}
\end{subfigure}
\begin{subfigure}[t]{0.01\textwidth}
\end{subfigure}
\begin{subfigure}[t]{0.48\textwidth}
    \centering
    \includegraphics[width=\linewidth,valign=m]{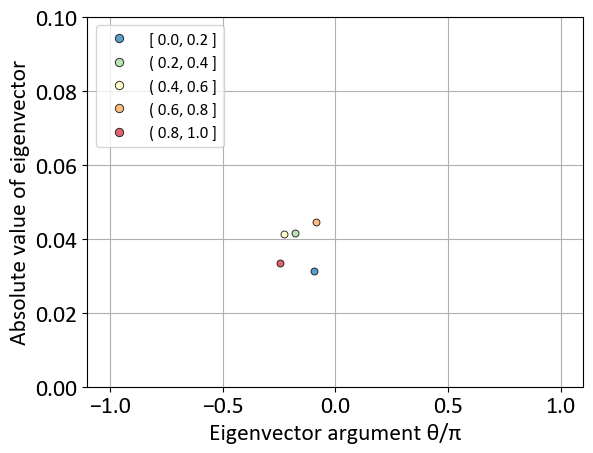}
    \caption{2022}
\end{subfigure}
\hfill

\caption{Barycentres for five population groups based on rank. The abscissa corresponds to the argument, and the ordinate corresponds to the absolute value (amplitude) of the eigenvector.
Panels (a-d) represent the entire period, 2020, 2021, and 2022, respectively. Note that time progresses from right to left.
}
\label{fig:eigen1populationbary}
\end{figure}

To see this trend more clearly,
we calculate the barycentres.
Since population is not a categorical value, it cannot be directly used to group elements as is done with the regions.
Therefore, the elements are divided into five levels based on rank,
and the barycentres are calculated for each level (Figure \ref{fig:eigen1populationbary}).

In terms of the lead and lag relationships, no obvious trends are observed.
Since it is natural to assume that infections in countries with large populations are more likely to spread to other countries,
this result is counterintuitive.
This finding suggests that causality cannot be explained by the population alone.
Regarding the mean distances from the barycentres,
we observe a tendency
similar to the regional analyses; i.e., the distances are greater in 2021 and 2022 than in 2020 (SI Table 2).

Since not only population but also wealth indicates high mobility and therefore the possibility of infection in other countries,
we also examine the countries by GDP per capita.
In contrast to the population results, the GDP per capita results clearly show causality.
The results indicate that countries with higher GDP per capita show leading tendencies and higher amplitudes in 2020 (SI Figure 9), which can be clearly observed in the barycentres (SI Figure 10).
However, this trend is not observed in 2021 or 2022. Instead, the opposite trend is observed, i.e., countries with low GDP per capita lead to more infections in other countries, although the amplitude is not considerably larger than the amplitude in those same countries in 2020.

As a summary of population and GDP characterizations,
population explains only the amplitude in 2020, 
but GDP per capita explains both the amplitude and causality in 2020.
However, in 2021 and 2022, population and GDP per capita alone cannot explain the amplitude or causality, let alone demonstrate the contradictory causality; i.e. low GDP per capita leads to infection.

\subsection*{Trends and transitions: interpretation by countermeasures and vaccination rates}

Countermeasures and nonpharmaceutical interventions (NPIs) can be implemented to reduce the spread of disease.
Therefore, 
we investigate the effectiveness of these measures by using the stringency index proposed by Our World in Data \cite{ourworldindata23}.
Intuitively, if stringent measures are imposed, the countries should be located at a lower amplitude than countries without measures and should have lagging positions.
However, the stringency index results do not show any prominent trends over either the entire period or in the yearly analyses (SI Figure 11 for the raw plots and SI Figure 12 for the barycentres)).
Notably, the order of the amplitude seems to have a negative correlation with the index.

We also examine the containment and health index, which is also
proposed by Our World in Data \cite{ourworldindata23}.
The index was developed based on the stringency index and considers advanced prevention measures.
The results of the index do not differ considerably from the stringency index results 
(SI Figure 13 for the raw plots and SI Figure 14 for the barycentres).
However, in 2022, the order of the amplitude is related to the index;
i.e., strong measures lead to smaller amplitudes. 

Like stringent countermeasures, the vaccination rate should delay the spread of infections and can be promoted by governments.
Therefore, we examine the eigenvectors with respect to the vaccination rate.
In contrast to the stringency index results,
the vaccination rate results reveal the effect of the delay, especially in 2022
(SI Figure 15 for the raw plots and SI Figure 16 for the barycentres).
Since vaccines were available to only a limited number of countries and were not adequately supplied to all in 2020, the eigenvector cannot be evaluated for this period. On the other hand, although the vaccine was rolled out in 2021, we cannot observe an effect from the delay of vaccination.

As a summary of this subsection,
the results of analyses based on the stringency and the containment and health indexes may be counterintuitive in the sense of causality
because we expect that implementing prevention measures would delay the spread of infections.
However, we observe that the amplitude is suppressed by countermeasures only in 2022, suggesting that countermeasures take time to be effective.
Regarding vaccination,
the results show clear causality consistent with the increased vaccination rate in 2022, which cannot be observed with other characterizations thus far, especially in the analysis by countermeasures.

\subsection*{Trends and transitions: interpretation by democracy}

Finally, we consider the degree of democracy in the countries.
We investigate this factor because imposing countermeasures may lead to resistance, especially in democratic countries.
Therefore, we examine the eigenvectors by the democracy index.

\begin{figure}[ht]
\centering

\begin{subfigure}[t]{0.48\textwidth}
    \centering
    \includegraphics[width=\linewidth,valign=m]{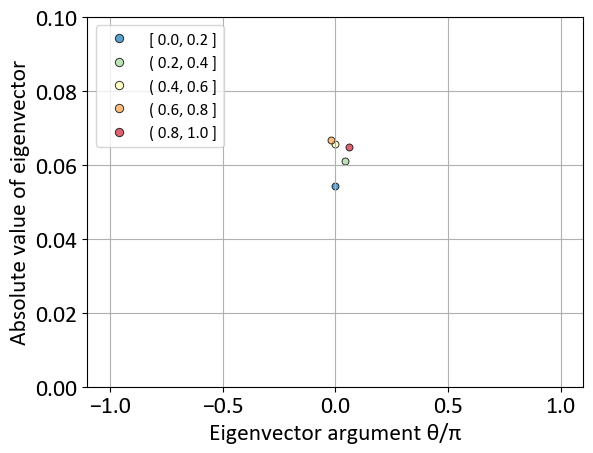}
    \caption{Entire period} 
\end{subfigure}
\begin{subfigure}[t]{0.01\textwidth}
\end{subfigure}
\begin{subfigure}[t]{0.48\textwidth}
    \centering
    \includegraphics[width=\linewidth,valign=m]{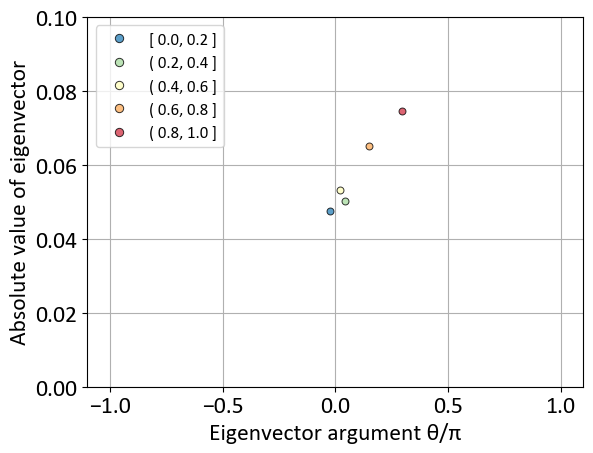}
    \caption{2020} 
\end{subfigure}
\hfill

\vspace{3ex}

\begin{subfigure}[t]{0.48\textwidth}
    \centering
    \includegraphics[width=\linewidth,valign=m]{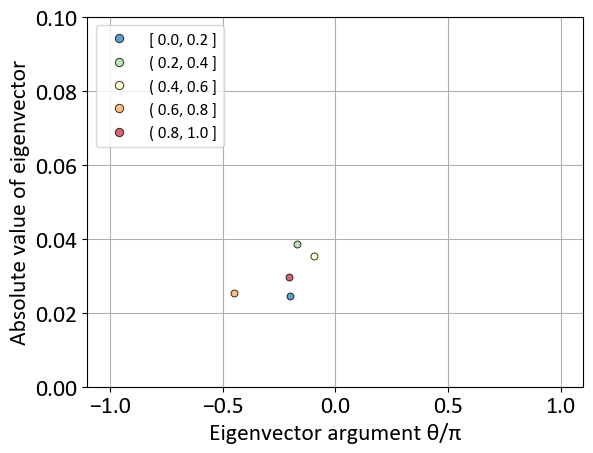}
    \caption{2021}
\end{subfigure}
\begin{subfigure}[t]{0.01\textwidth}
\end{subfigure}
\begin{subfigure}[t]{0.48\textwidth}
    \centering
    \includegraphics[width=\linewidth,valign=m]{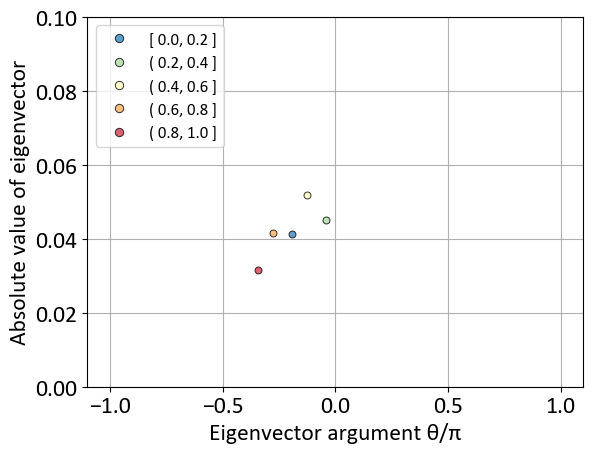}
    \caption{2022}
\end{subfigure}
\hfill

\caption{Barycentres for each of five groups by democracy index based on rank. The abscissa corresponds to the argument, and the ordinate corresponds to the absolute value (amplitude) of the eigenvector.
Panels (a-d) represent the entire period, 2020, 2021, and 2022, respectively. Note that time progresses from right to left.
}
\label{fig:eigen1demobary}
\end{figure}

The results show no observable trend over the entire period (Figure \ref{fig:eigen1demobary} for the barycentres and SI Figure 17 for the raw plots).
However, 
there are clear yearly changes in the democracy index.
Democratic countries take a leading position and have larger amplitudes in 2020, shifting to a middle position in 2021. Then, they take a lagging position and have smaller amplitudes in 2022, which is clearly observed in the barycentres.
This means that the trends in nondemocratic countries reversed during these years.

In summary, 
the democracy index shows clear interpretable results
unlike the other characterizations.
This means that other characterizations can be used to interpret the trend in some years but not all years.
However, the democracy index shows clear interpretable trends, especially in 2020 and 2022, and these trends completely reverse over the course of the whole period.

\subsection*{The 2nd and 3rd ranked eigenvectors}

We have discussed the first eigenvectors of the entire period and each year thus far.
For the entire period,
three eigenvalues are significant over the entire period.
Although the third eigenvalue is close to the significance interval,
the second eigenvector clearly indicates significance.
Therefore, it is worth examining the second eigenvector as well as the first eigenvector.

The second-rank eigenvector of the entire period can be described by the presence of strong lead and lag over the entire period (SI Figure 18 for the raw plots and SI Figure 19 for the barycentres), which means that the eigenvector spreads over the three years. On the other hand, the first eigenvector only spread over approximately one year (SI Figure 6 etc.).
Additionally, in each interpretation, i.e., the region, population, stringency index, and democracy results, the lead and lag become more pronounced, while no lead or lag relationships are clearly observed with the first eigenvector.

The above findings for the entire period are almost consistent with the analysis of the first eigenvector based on 2020 data (SI Figure 20 for the raw plots and SI Figure 21 for the barycentres).
In other words, the second eigenvector over the entire period captures the trends observed in the 2020 data.
In addition, unsurprisingly, the third eigenvector over the entire period is aligned with the first eigenvectors in the 2021 and 2022 data (SI Figures 26 for the barycentres of the entire period).
The details of the analyses are described in 
SI Section 3.

\subsection*{Discussion}

The purpose of this study is
to reveal patterns of international interaction in the spread of infection
including not only the magnitude but also the lead and lag times.
Then, we examine the interaction patterns 
from the perspective of regions, population, restrictions and countermeasures, vaccination rates, and government.
As an overall result,
we find that 
there are two primary trends occurring in 2020 and 2021--2022
that interchange between each other.
In addition, 
none of the perspectives showed the consistent lead and lag positions.
Based on the above analysis results,
we present some synthesized interpretations of the findings.

First, the shapes of the plots in the CHPCA results differ between 2020 and 2021--2022.
The large-amplitude countries do not show substantial lead or lag relationships with other locations in 2020, i.e., the plots show triangular shapes. However, in later years, the shape changes, spreading horizontally. In addition, all perspectives used for the analyses, i.e., region, population, GDP per capita, stringency of the countermeasures, vaccination rates, and government, show different trends between 2020 and 2021--2022.
This fact is intriguing and can probably be explained by
social adaptations.
In 2020, when the largest pandemic in modern history first impacted society,
people in different countries did not know how to respond to this tremendous threat and could not anticipate the outcomes of their countermeasures.
As a consequence, countries with characteristics that typically cause increase infections, i.e., larger populations, higher GDP per capita (or high mobility), and high degree of democracy (or freedom of movement), show leading and high-impact positions.
Since any countermeasures to prevent infections incur economic and social losses,
infection prevention measures, and economic and societal impacts must be balanced.
Governments need elaborate strategies that consider each country's situation, which take time to develop.
Overall, diverse infection prevention strategies were developed in that first year,
and the trends observed in 2020 disappeared, with new trends emerging.

Second, corresponding to the above analyses, in the regional analysis,
European countries showed leading trends and had higher impacts in 2020, 
while African and American countries showed leading trends in 2021 and 2022, respectively, yet there were no high-impact countries in those years.
In 2020, countermeasures had not yet been implemented, and high-mobility countries, i.e., European countries, showed relatively leading trends.
In recent years, due to widespread adoption of the countermeasures, advanced countries showed lagging positions, while Africa and the Americas showed leading positions.

Third, populous countries showed leading trends and had greater impacts in 2020, which is intuitive as a result because effective countermeasures were not applied in other countries.
However, in the later period, the population results did not follow this trend.
Although the cause of this difference is presumed to be the countermeasures described above, this finding is surprising simply because more populous countries can be expected to cause the spread of infections.
Therefore, this result indicates that if countermeasures to prevent infection are properly implemented,
the spread of infections from larger countries can be mitigated, even if the country has a large population.
Moreover, through an examination focusing on GDP per capita,
we found that countries with higher GDP per capita have leading and high-amplitude positions in 2020 but that they have lagging and low-amplitude positions in 2021--2022.
This result can be explained by the increased mobility of the population and the ability of the government to impose effective countermeasures.
Since wealthier countries have higher international travel rates, they have a greater risk of rapidly increasng the spread of infections.
However, since the governments of these countries can 
afford the cost of implementing more countermeasures, they can delay and mitigate the spread of infection.

Fourth, along with the above discussions,
countermeasures are more effective in later years.
This can be observed by the reduced infection rates associated with more stringent countermeasures.
Moreover, the effect is more evident for stricter countermeasures.
In addition, the vaccination rate can reduce the impact of infections, with a particularly pronounced effect on delayed infections.
These results show that it takes approximately one year for each country to prepare effective countermeasures, and the countermeasures change the trends in the CHPCA results.

Finally, regarding results of analysis broken down by the democracy index,
a very clear trend can be observed in 2020, with more democratic countries exhibiting higher amplitudes and leading positions.
This trend reversed in 2021--2022.
The results may reflect not only that
democratic countries have difficulty imposing stricter measures 
but also that the number of infections was not censored at the beginning of the pandemic.
After the disease spread widely and became common, nondemocratic countries could be expected to report an accurate number of infections.

\section*{Conclusion}

We analysed global infection case data from January 3, 2020, to the end of 2022, as well as the separate yearly data, 
and applied CHPCA to 
comprehensively elucidate international interactions with not only the impact but the lead and lag.
Importantly, CHPCA is similar to PCA; however, CHPCA enables the investigation of relationships not only at the same point in time but also with lead and lag times.
Similar to PCA, the eigenvectors in CHPCA should be interpreted from various points of view.
Therefore, we separated countries by region, population, GDP, stringency of the countermeasures, vaccination rates, and democratic governments.

The CHPCA results reveal that there are dominant eigenvectors and one or two additional eigenvectors.
Although the latter eigenvectors are statistically significant,
since the first eigenvectors are dominant,
we investigated mainly the first eigenvectors.
It is evident that during the period of 2020--2022, two primary trends coexist, with different trends dominating in 2020 and 2021--2022 and they interchange between each other, which can be observed from the analyses of the entire and yearly periods.
None of the viewpoints, i.e., region, population, GDP per capita, stringency of the countermeasures, vaccination rate, and democratic degree, can fully explain the trends.
By examining these viewpoints, it can be concluded that during the early stages of the pandemic, activities that facilitated infections, such as intercountry travel and geographical distance, had significant temporal influences on the spread of infection.
However, as societies have implemented interventions,
the countermeasures had positive impacts on mitigating these dynamics,
and the spread of the infection became more complex, as captured by the analyses performed in this paper.

The study results cannot be obtained by other methods, such as conventional PCA or other regression-based methods because no other methods can be used to comprehensively extract the unseen causality in multivariate data. Therefore, the results of this study have a unique value.

The study results have several implications for preparations to prevent the spread of infection.
The pandemic started mainly in Europe, which includes many countries with high GDP and democratic regimes, i.e., high-mobility countries.
However, countermeasures drastically changed the trend in disease spread.
Moreover, less economically prosperous countries
could become sources of infections over time.
Since some countries cannot manage pandemics ,
we suggest that international efforts to promote measures in developing countries may contribute to pandemic containment, as indicated by the findings in this paper.

In terms of the methodology used in this paper,
CHPCA shows promising potential for extracting
obscured but interesting causality from large scale data.
The typical cases where CHPCA may be applied are normally in research on societal issues, such as consumption patterns or prices of goods, because goods have substitutability and they may have unseen causality.

\bibliography{ref}

\section*{Acknowledgements}

The authors thank Yuito Nishi for his technical contribution.

\section*{Author contributions statement}

H.I. conceived the project, conducted the analyses, and wrote the manuscript. H.I., W.S., and Y.F. analysed the results. All the authors reviewed the manuscript. 




\clearpage

\clearpage

\clearpage

\clearpage

\clearpage

\end{document}


\fontsize{13pt}{15pt}\selectfont
\begin{center}
{\bf The Coexistence of Infection Spread Patterns in the Global Dynamics of COVID-19 Dissemination}
\vspace{2ex}

{\bf Supplementary Information}
\vspace{2ex}

\fontsize{11pt}{15pt}\selectfont
Hiroyasu Inoue\\
Wataru Souma\\
Yoshi Fujiwara
\vspace{3ex}

\end{center}
\fontsize{10pt}{12pt}\selectfont

\section{Supplementary figures}

\begin{figure}[ht]
\centering
\includegraphics[width=.6\linewidth]{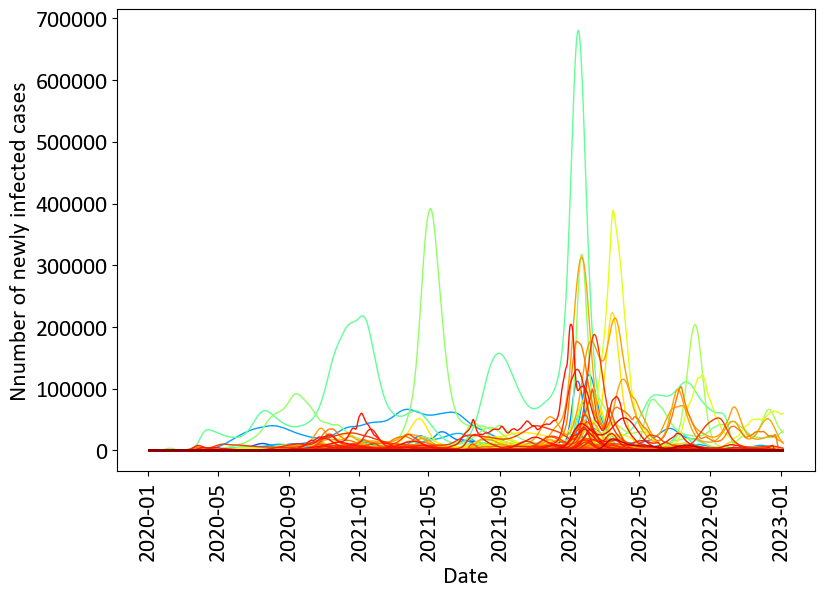}
\caption{Number of infected cases in all countries. Each colour indicates a country. The correspondence of the colours with the countries is shown in SI Figure \ref{fig:legend}.}
\label{fig:numberofinfected}
\end{figure}

\begin{figure}[ht]
\centering
\includegraphics[width=.6\linewidth]{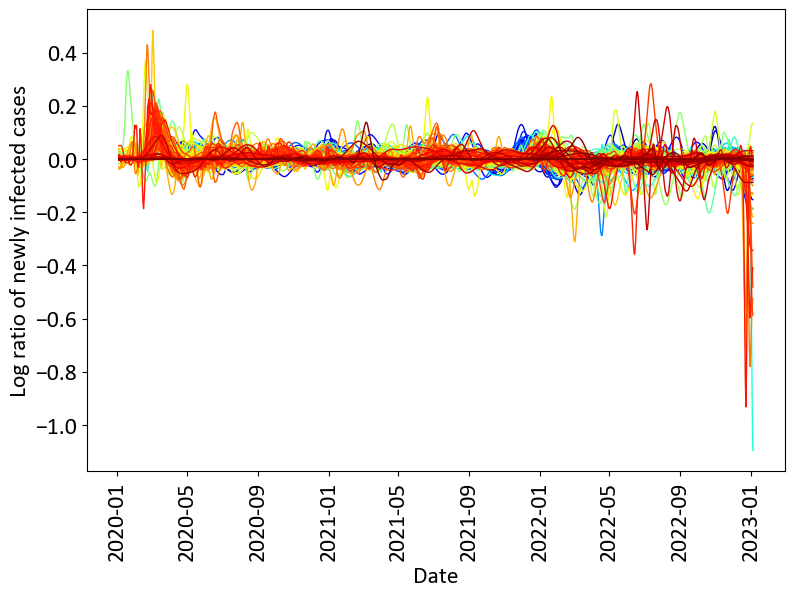}
\caption{Log ratio of infected cases for all countries. Each colour indicates a country. The correspondence of the colours with the countries is shown in SI Figure \ref{fig:legend}.}
\label{fig:logratio}
\end{figure}

\begin{figure}[ht]
\centering
\includegraphics[width=\linewidth]{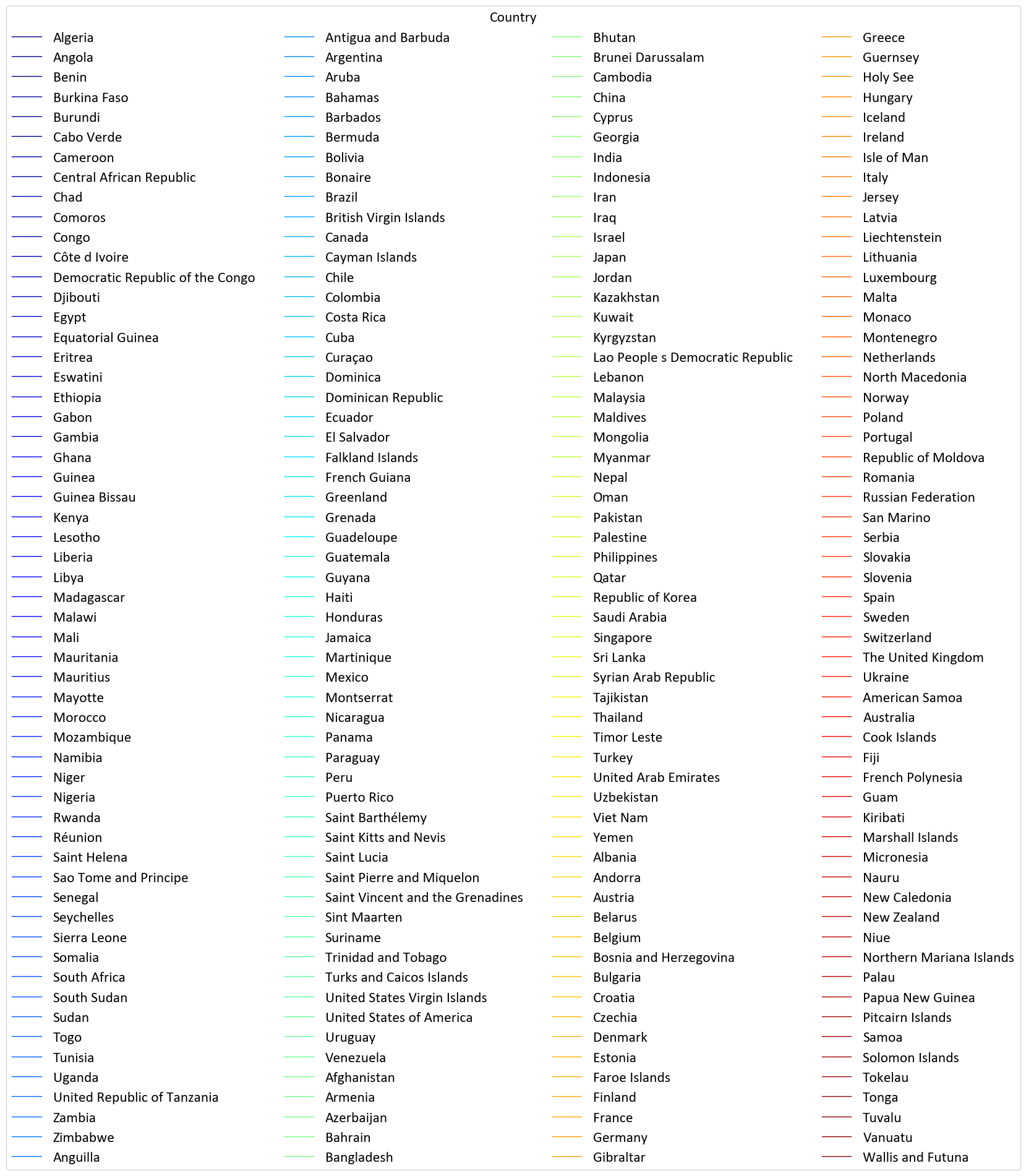}
\caption{Correspondence of colours with countries. These colours are used in SI Figures \ref{fig:numberofinfected} and \ref{fig:logratio}.}
\label{fig:legend}
\end{figure}

\begin{figure}[ht]
\centering
\includegraphics[width=\linewidth]{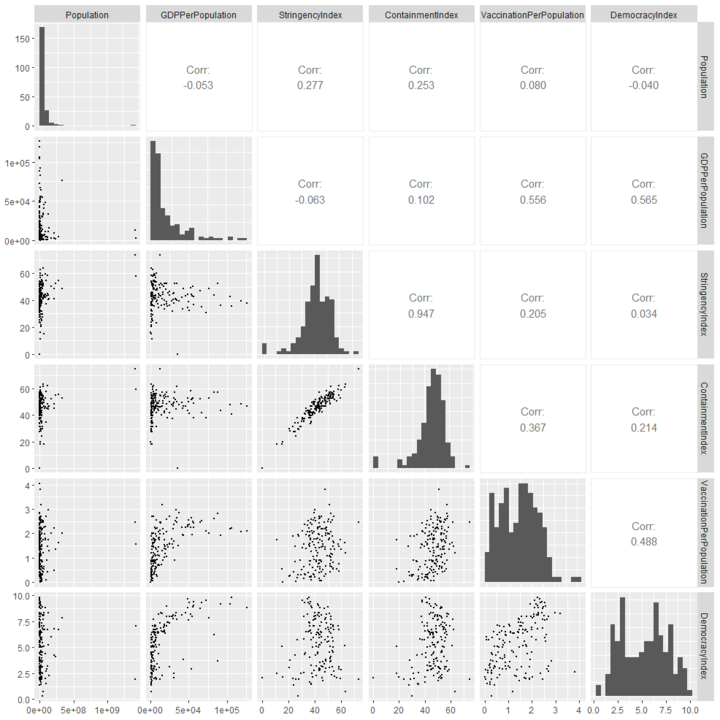}
\caption{Pair plots between auxiliary data points.}
\label{fig:aux}
\end{figure}

\begin{figure}[ht]
\centering

\begin{subfigure}[t]{0.45\textwidth}
    \centering
    \includegraphics[width=.9\linewidth]{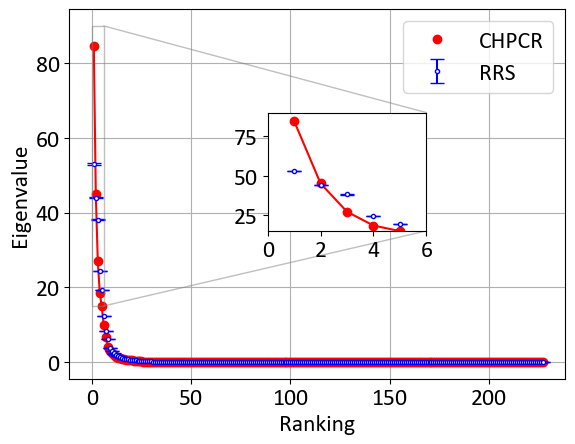}
    \caption{2020} 
\end{subfigure}
\begin{subfigure}[t]{0.45\textwidth}
    \centering
    \includegraphics[width=.9\linewidth]{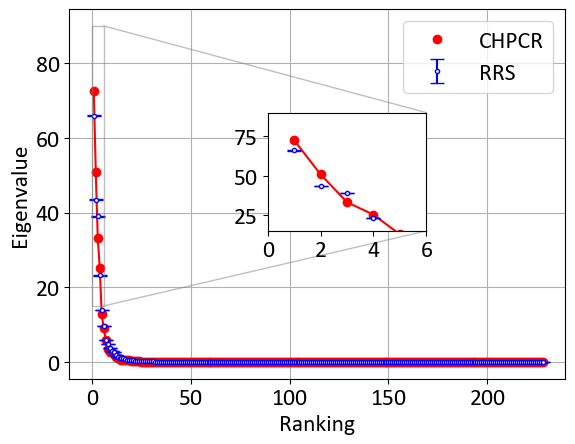}
    \caption{2021}
\end{subfigure}
\hfill

\vspace{3ex}

\begin{subfigure}[t]{0.45\textwidth}
    \centering
    \includegraphics[width=.9\linewidth]{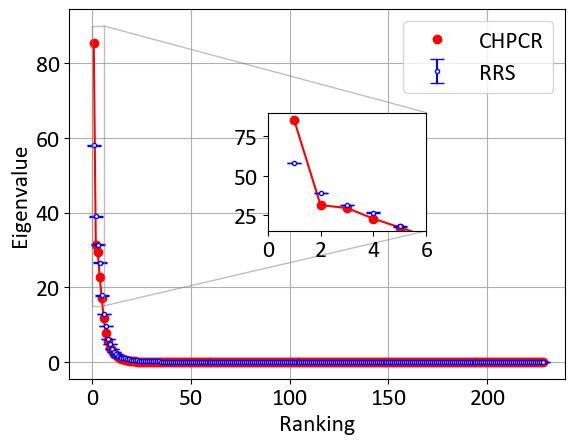}
    \caption{2022}
\end{subfigure}
\begin{subfigure}[t]{0.45\textwidth}
\hfill
\end{subfigure}
\hfill

\caption{Scree plots of CHPCA and the RRS. Each panel covers different years, i.e., 2020, 2021, and 2020. The error bars for the RRS indicate the 1\% significance level, which is 2.33 times the standard error.}
\label{fig:scree}
\end{figure}

\begin{figure}[ht]
\centering

\begin{subfigure}[t]{0.48\textwidth}
    \centering
    \includegraphics[width=\linewidth,valign=m]{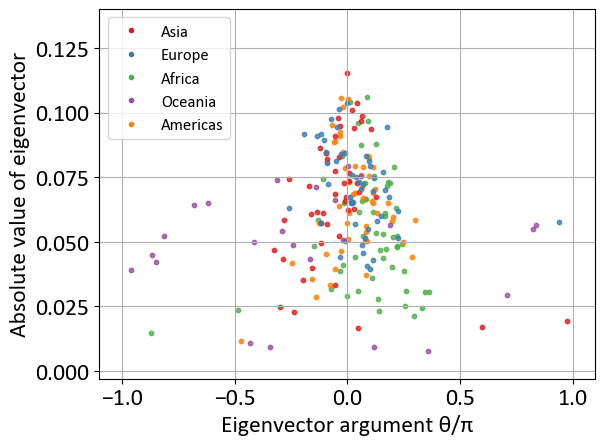}
    \caption{Entire period} 
\end{subfigure}
\begin{subfigure}[t]{0.01\textwidth}
\end{subfigure}
\begin{subfigure}[t]{0.48\textwidth}
    \centering
    \includegraphics[width=\linewidth,valign=m]{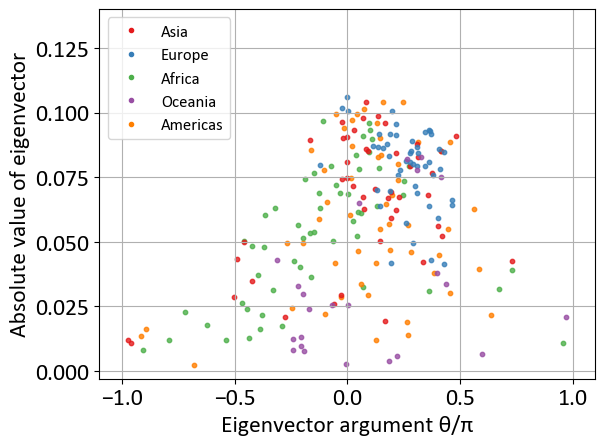}
    \caption{2020} 
\end{subfigure}
\hfill

\vspace{3ex}

\begin{subfigure}[t]{0.48\textwidth}
    \centering
    \includegraphics[width=\linewidth,valign=m]{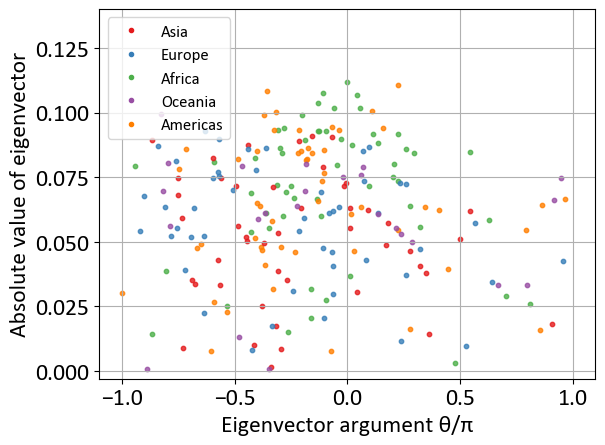}
    \caption{2021}
\end{subfigure}
\begin{subfigure}[t]{0.01\textwidth}
\end{subfigure}
\begin{subfigure}[t]{0.48\textwidth}
    \centering
    \includegraphics[width=\linewidth,valign=m]{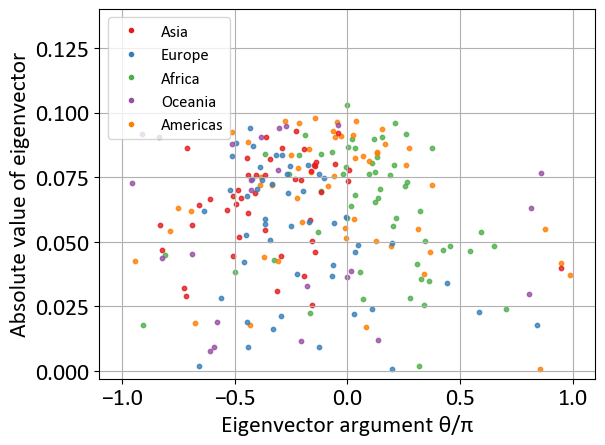}
    \caption{2022}
\end{subfigure}
\hfill

\caption{Eigenvectors with the first eigenvalue coloured by region. The abscissa corresponds to the argument, and the ordinate corresponds to the absolute value (amplitude) of the eigenvector. Note that time progresses from right to left. Panels (a-d) show the entire period, 2020, 2021, and 2022, respectively.}
\label{fig:eigen1region}
\end{figure}

\begin{figure}[ht]
\centering

\begin{subfigure}[t]{0.48\textwidth}
    \centering
    \includegraphics[width=\linewidth,valign=m]{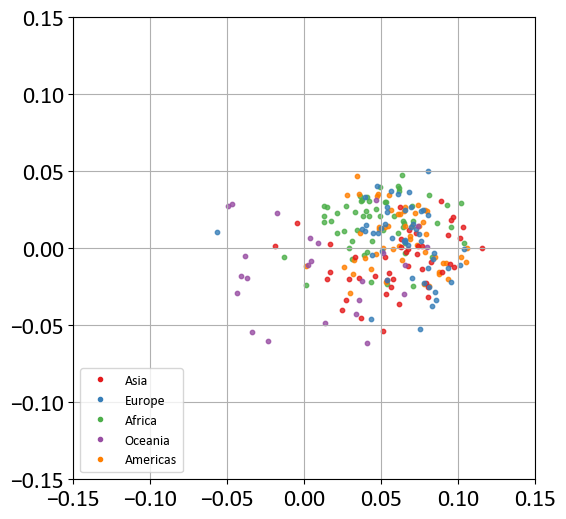}
    \caption{Entire period} 
\end{subfigure}
\begin{subfigure}[t]{0.01\textwidth}
\end{subfigure}
\begin{subfigure}[t]{0.48\textwidth}
    \centering
    \includegraphics[width=\linewidth,valign=m]{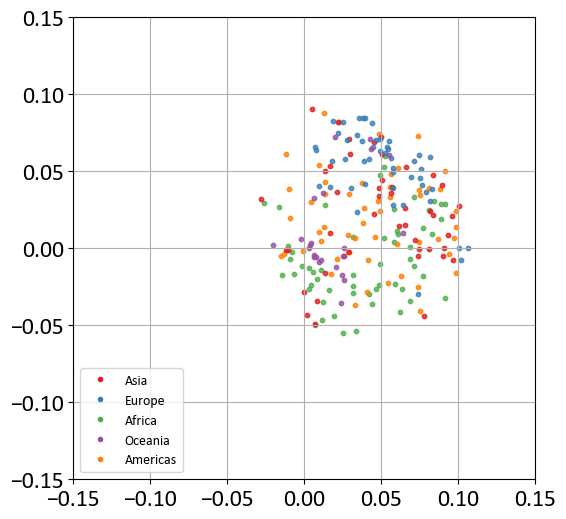}
    \caption{2020} 
\end{subfigure}
\hfill

\vspace{3ex}

\begin{subfigure}[t]{0.48\textwidth}
    \centering
    \includegraphics[width=\linewidth,valign=m]{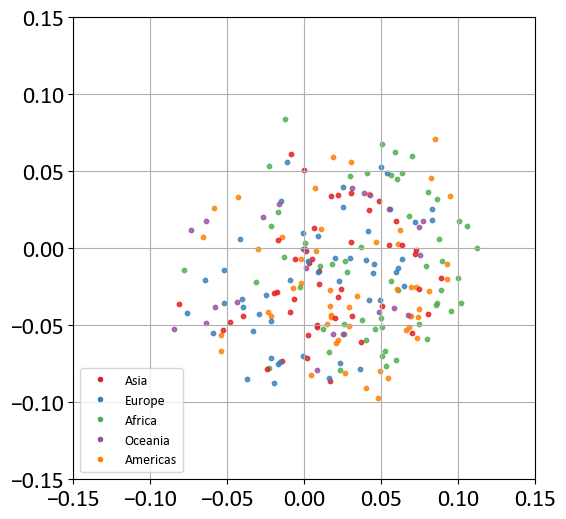}
    \caption{2021}
\end{subfigure}
\begin{subfigure}[t]{0.01\textwidth}
\end{subfigure}
\begin{subfigure}[t]{0.48\textwidth}
    \centering
    \includegraphics[width=\linewidth,valign=m]{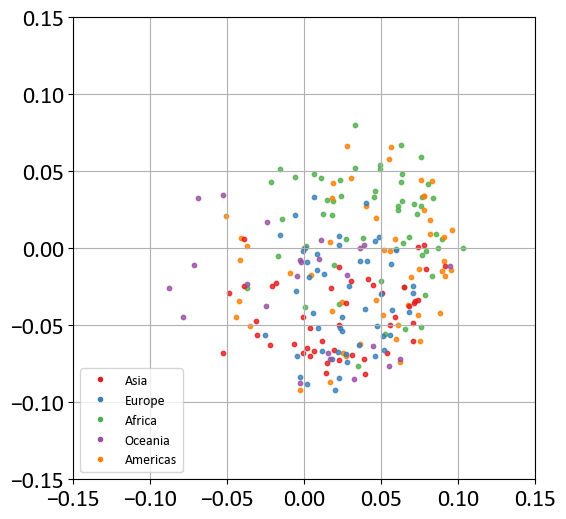}
    \caption{2022}
\end{subfigure}
\hfill

\caption{Eigenvectors, with the first eigenvalue coloured by region. The abscissa corresponds to the real axis, and the ordinate corresponds to the imaginary axis.
Panels (a-d) represent the entire period, 2020, 2021, and 2022, respectively. Note that time progresses from right to left.}
\label{fig:eigen1regioncomplex}
\end{figure}

\begin{figure}[ht]
\centering

\begin{subfigure}[t]{0.48\textwidth}
    \centering
    \includegraphics[width=\linewidth,valign=m]{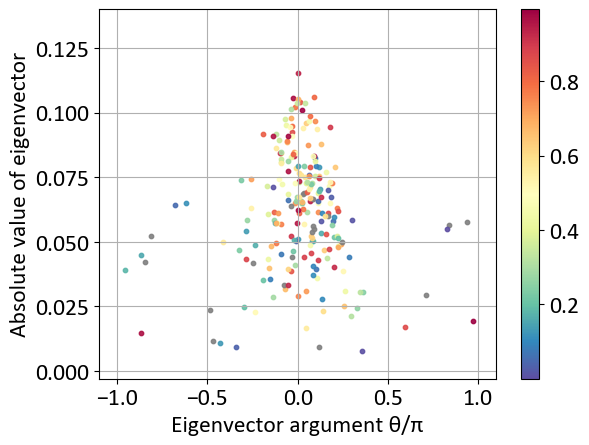}
    \caption{Entire period} 
\end{subfigure}
\begin{subfigure}[t]{0.01\textwidth}
\end{subfigure}
\begin{subfigure}[t]{0.48\textwidth}
    \centering
    \includegraphics[width=\linewidth,valign=m]{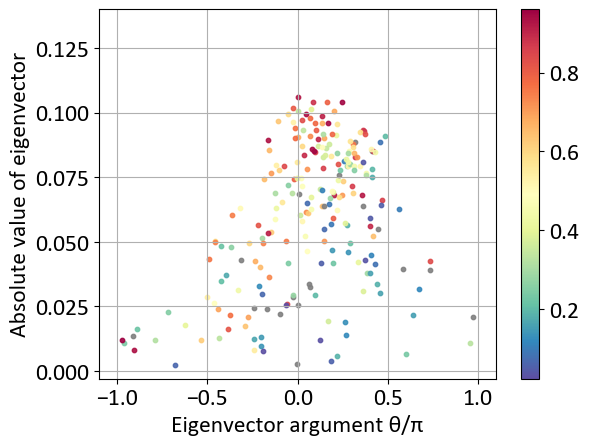}
    \caption{2020} 
\end{subfigure}
\hfill

\vspace{3ex}

\begin{subfigure}[t]{0.48\textwidth}
    \centering
    \includegraphics[width=\linewidth,valign=m]{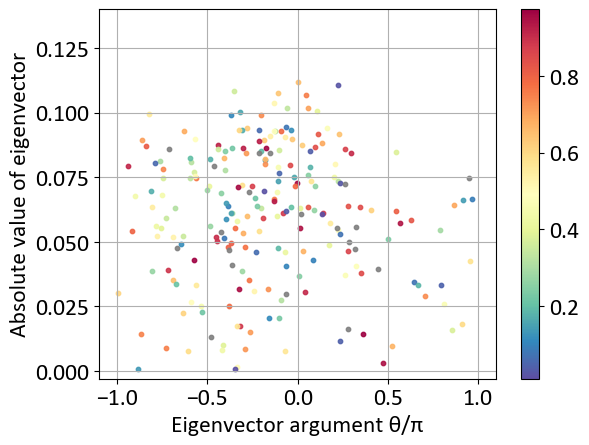}
    \caption{2021}
\end{subfigure}
\begin{subfigure}[t]{0.01\textwidth}
\end{subfigure}
\begin{subfigure}[t]{0.48\textwidth}
    \centering
    \includegraphics[width=\linewidth,valign=m]{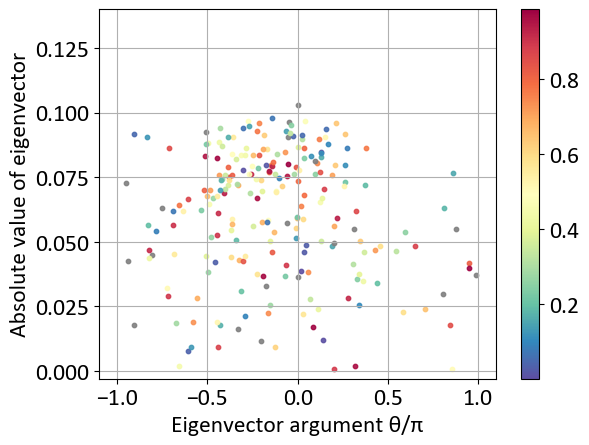}
    \caption{2022}
\end{subfigure}
\hfill

\caption{Eigenvectors with the first eigenvalue coloured according to population. The abscissa corresponds to the argument, and the ordinate corresponds to the absolute value (amplitude) of the eigenvector. Note that time progresses from right to left. Panels (a-d) represent the entire period, 2020, 2021, and 2022, respectively. The population data for each year are used for the analysis of the corresponding year, while the 2020 population data are used for the analysis of the entire period. In addition, the colour denotes the rank of the analysed countries by population. If a country has no population record, it is coloured grey.}
\label{fig:eigen1population}
\end{figure}

\begin{figure}[ht]
\centering

\begin{subfigure}[t]{0.48\textwidth}
    \centering
    \includegraphics[width=\linewidth,valign=m]{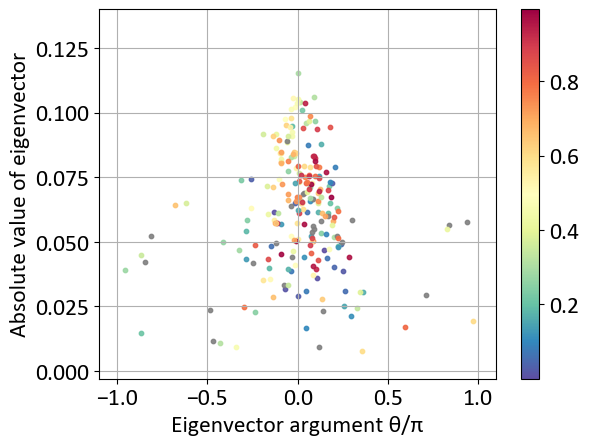}
    \caption{Entire period} 
\end{subfigure}
\begin{subfigure}[t]{0.01\textwidth}
\end{subfigure}
\begin{subfigure}[t]{0.48\textwidth}
    \centering
    \includegraphics[width=\linewidth,valign=m]{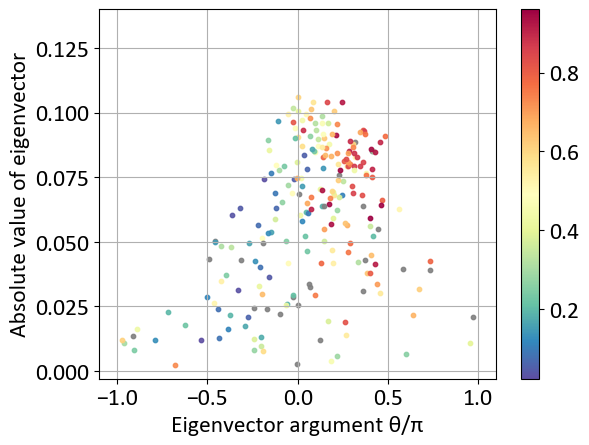}
    \caption{2020} 
\end{subfigure}
\hfill

\vspace{3ex}

\begin{subfigure}[t]{0.48\textwidth}
    \centering
    \includegraphics[width=\linewidth,valign=m]{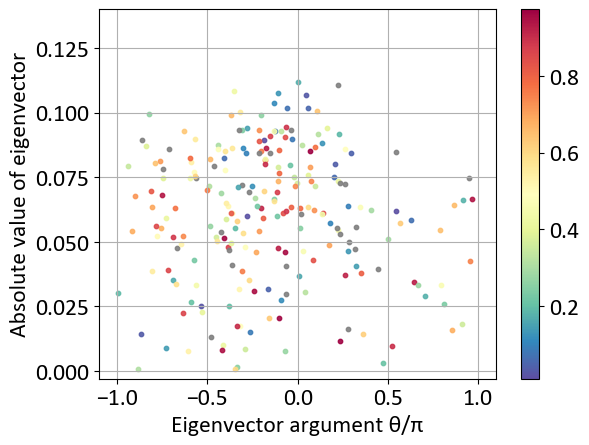}
    \caption{2021}
\end{subfigure}
\begin{subfigure}[t]{0.01\textwidth}
\end{subfigure}
\begin{subfigure}[t]{0.48\textwidth}
    \centering
    \includegraphics[width=\linewidth,valign=m]{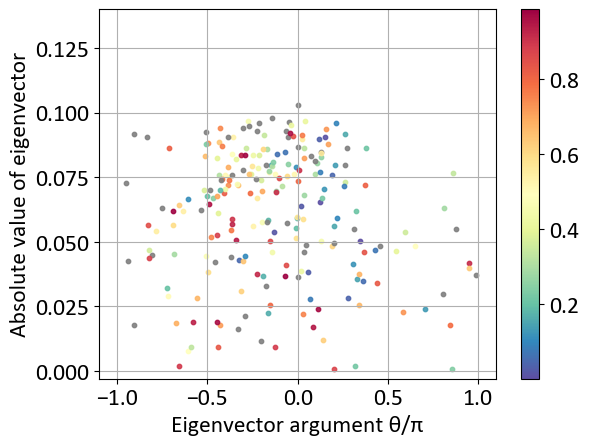}
    \caption{2022}
\end{subfigure}
\hfill

\caption{Eigenvectors, with the first eigenvalue coloured according to the GDP per capita. The abscissa corresponds to the real axis, and the ordinate corresponds to the imaginary axis. Panels (a-d) represent the entire period, 2020, 2021, and 2022, respectively. The population and GDP data for each year are used for the corresponding year, while the 2020 population and GDP data are used for the analysis of the entire period. In addition, the colour indicates the rank of the analysed countries by GDP per capita. If a country has no population or GDP record, it is coloured grey.}
\label{fig:eigen1gdpcapita}
\end{figure}

\begin{figure}[ht]
\centering

\begin{subfigure}[t]{0.48\textwidth}
    \centering
    \includegraphics[width=\linewidth,valign=m]{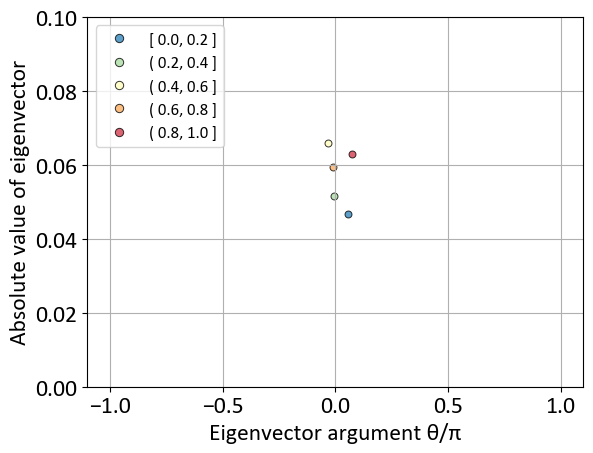}
    \caption{Entire period} 
\end{subfigure}
\begin{subfigure}[t]{0.01\textwidth}
\end{subfigure}
\begin{subfigure}[t]{0.48\textwidth}
    \centering
    \includegraphics[width=\linewidth,valign=m]{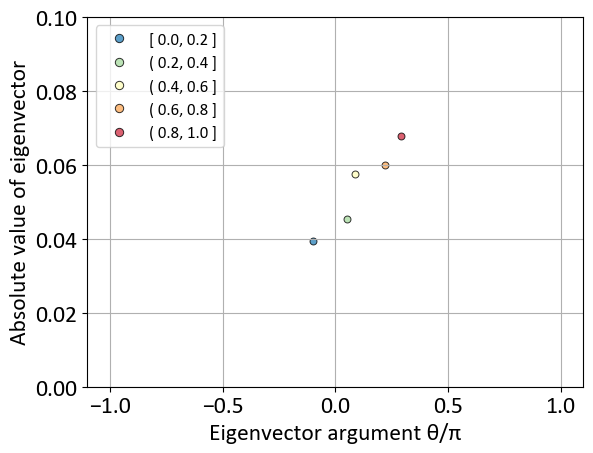}
    \caption{2020} 
\end{subfigure}
\hfill

\vspace{3ex}

\begin{subfigure}[t]{0.48\textwidth}
    \centering
    \includegraphics[width=\linewidth,valign=m]{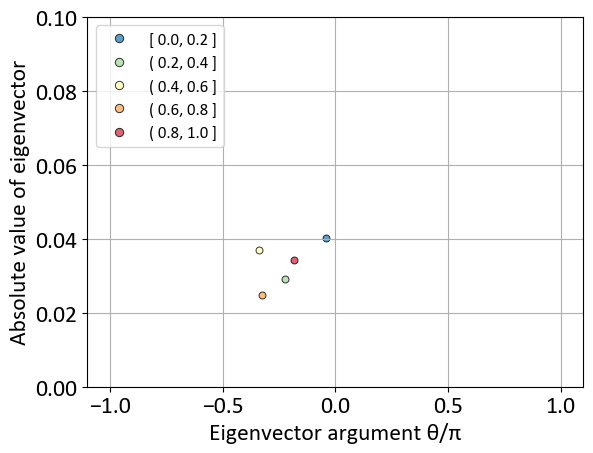}
    \caption{2021}
\end{subfigure}
\begin{subfigure}[t]{0.01\textwidth}
\end{subfigure}
\begin{subfigure}[t]{0.48\textwidth}
    \centering
    \includegraphics[width=\linewidth,valign=m]{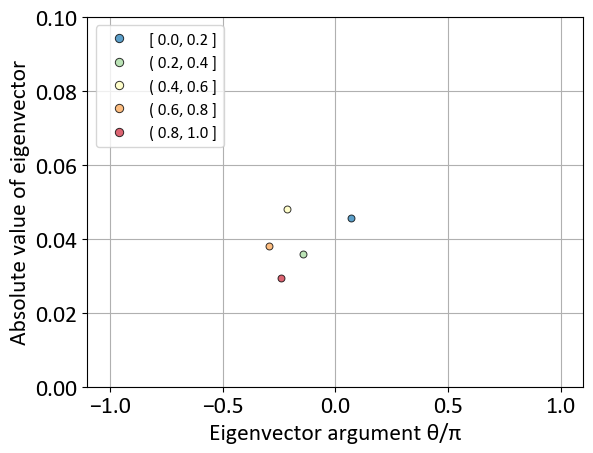}
    \caption{2022}
\end{subfigure}
\hfill

\caption{Barycentres for five groups of GDP per capita based on rank. The abscissa corresponds to the argument, and the ordinate corresponds to the absolute value (amplitude) of the eigenvector.
Panels (a-d) represent the entire period, 2020, 2021, and 2022, respectively. Note that time progresses from right to left.
}
\label{fig:eigen1gdpcapitabary}
\end{figure}

\begin{figure}[ht]
\centering

\begin{subfigure}[t]{0.48\textwidth}
    \centering
    \includegraphics[width=\linewidth,valign=m]{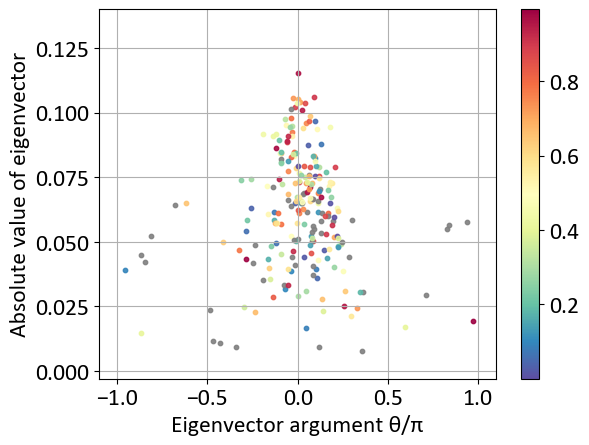}
    \caption{Entire period} 
\end{subfigure}
\begin{subfigure}[t]{0.01\textwidth}
\end{subfigure}
\begin{subfigure}[t]{0.48\textwidth}
    \centering
    \includegraphics[width=\linewidth,valign=m]{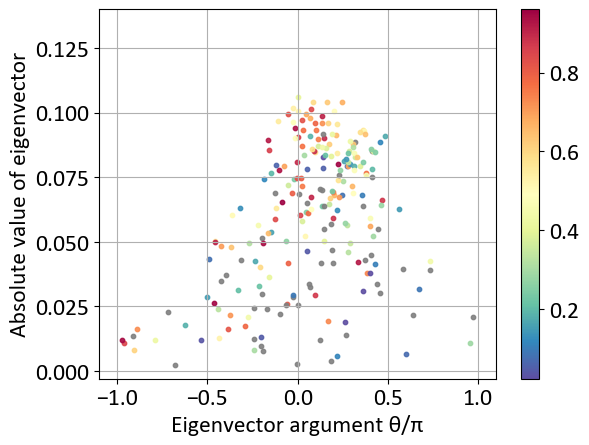}
    \caption{2020} 
\end{subfigure}
\hfill

\vspace{3ex}

\begin{subfigure}[t]{0.48\textwidth}
    \centering
    \includegraphics[width=\linewidth,valign=m]{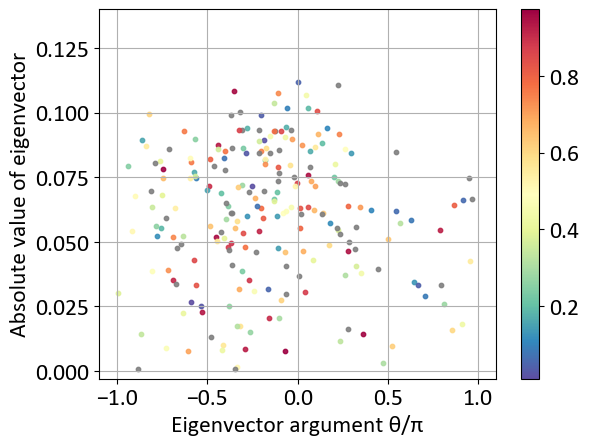}
    \caption{2021}
\end{subfigure}
\begin{subfigure}[t]{0.01\textwidth}
\end{subfigure}
\begin{subfigure}[t]{0.48\textwidth}
    \centering
    \includegraphics[width=\linewidth,valign=m]{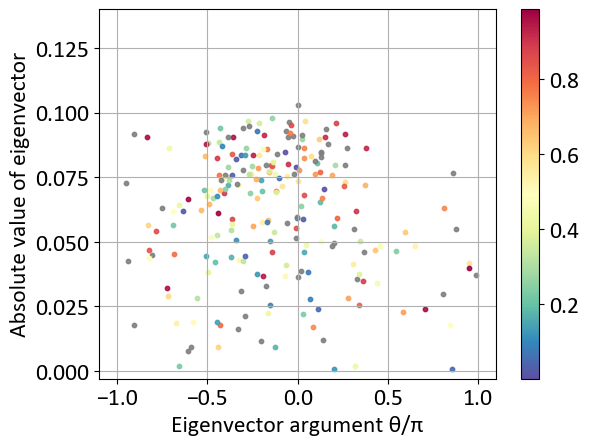}
    \caption{2022}
\end{subfigure}
\hfill

\caption{Eigenvectors, with the first eigenvalue coloured according to  the stringency index. The abscissa corresponds to the argument, and the ordinate corresponds to the absolute value (amplitude) of the eigenvector. Note that time progresses from right to left. Panels (a-d) represent the entire period, 2020, 2021, and 2022, respectively. The mean of each year's stringency index data is used for the corresponding year, while the mean of the entire period is used for the analysis of the entire period. In addition, the colour indicates the rank of the countries by stringency index. If a country has no stringency index record, it is coloured grey.}
\label{fig:eigen1stringency}
\end{figure}

\begin{figure}[ht]
\centering

\begin{subfigure}[t]{0.48\textwidth}
    \centering
    \includegraphics[width=\linewidth,valign=m]{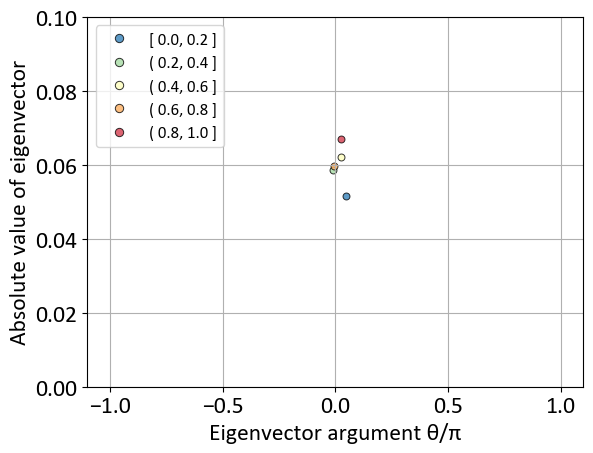}
    \caption{Entire period} 
\end{subfigure}
\begin{subfigure}[t]{0.01\textwidth}
\end{subfigure}
\begin{subfigure}[t]{0.48\textwidth}
    \centering
    \includegraphics[width=\linewidth,valign=m]{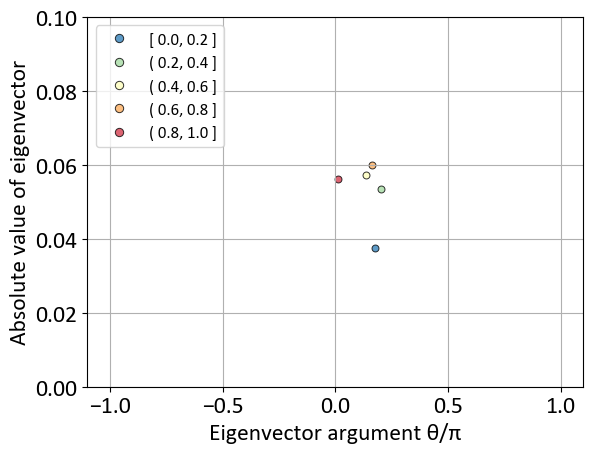}
    \caption{2020} 
\end{subfigure}
\hfill

\vspace{3ex}

\begin{subfigure}[t]{0.48\textwidth}
    \centering
    \includegraphics[width=\linewidth,valign=m]{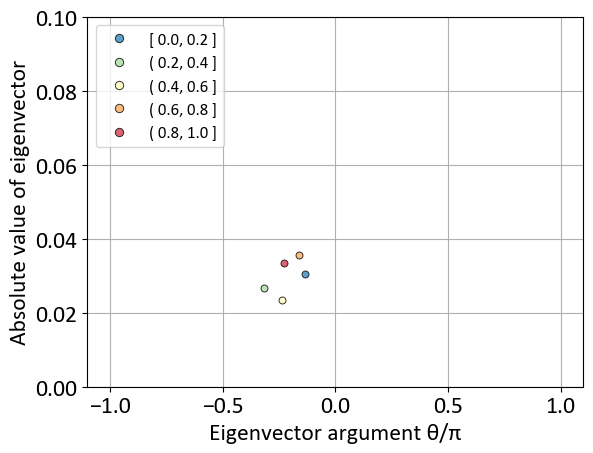}
    \caption{2021}
\end{subfigure}
\begin{subfigure}[t]{0.01\textwidth}
\end{subfigure}
\begin{subfigure}[t]{0.48\textwidth}
    \centering
    \includegraphics[width=\linewidth,valign=m]{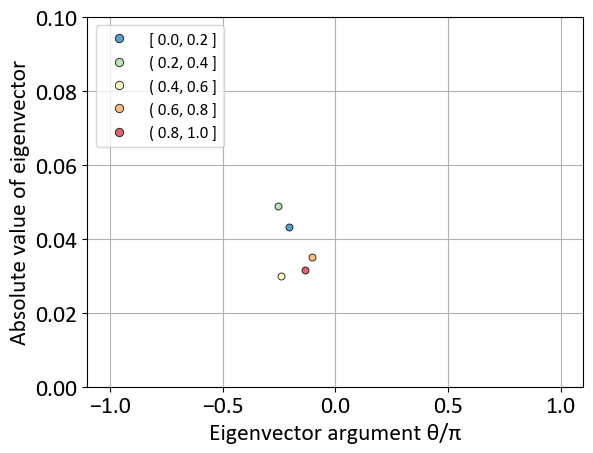}
    \caption{2022}
\end{subfigure}
\hfill

\caption{Barycentres for five groups of each stringency index based on rank. The abscissa corresponds to the argument, and the ordinate corresponds to the absolute value (amplitude) of the eigenvector.
Panels (a-d) represent the entire period, 2020, 2021, and 2022, respectively. Note that time progresses from right to left.
}
\label{fig:eigen1stringencybary}
\end{figure}

\begin{figure}[ht]
\centering

\begin{subfigure}[t]{0.48\textwidth}
    \centering
    \includegraphics[width=\linewidth,valign=m]{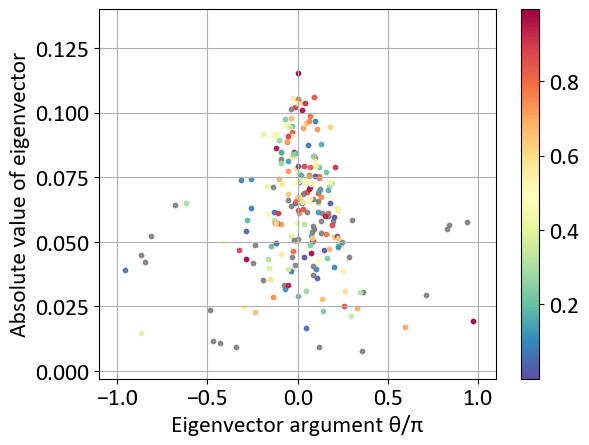}
    \caption{Entire period} 
\end{subfigure}
\begin{subfigure}[t]{0.01\textwidth}
\end{subfigure}
\begin{subfigure}[t]{0.48\textwidth}
    \centering
    \includegraphics[width=\linewidth,valign=m]{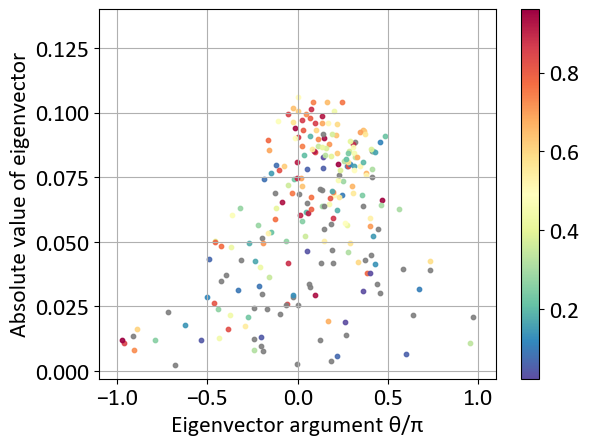}
    \caption{2020} 
\end{subfigure}
\hfill

\vspace{3ex}

\begin{subfigure}[t]{0.48\textwidth}
    \centering
    \includegraphics[width=\linewidth,valign=m]{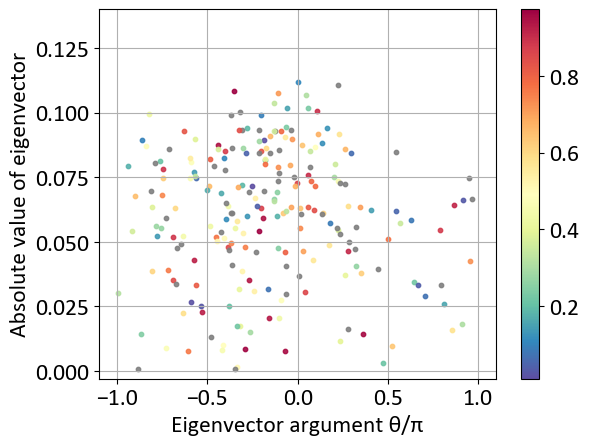}
    \caption{2021}
\end{subfigure}
\begin{subfigure}[t]{0.01\textwidth}
\end{subfigure}
\begin{subfigure}[t]{0.48\textwidth}
    \centering
    \includegraphics[width=\linewidth,valign=m]{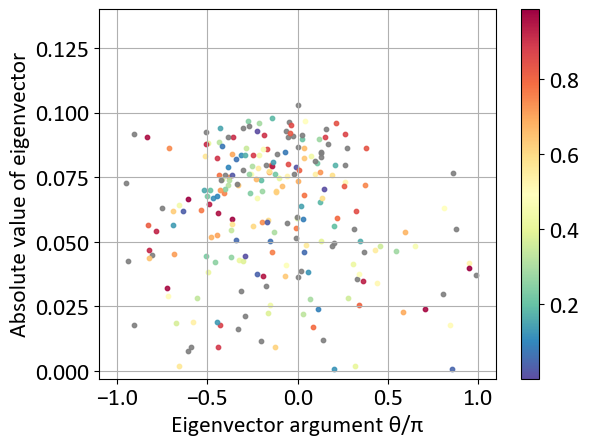}
    \caption{2022}
\end{subfigure}
\hfill

\caption{Eigenvectors, with the first eigenvalue coloured according to  the containment and health index. The abscissa corresponds to the argument, and the ordinate corresponds to the absolute value (amplitude) of the eigenvector. Note that time progresses from right to left. Panels (a-d) represent the entire period, 2020, 2021, and 2022, respectively. The mean of each year's stringency index data is used for the corresponding year, while the mean of the entire period is used for the entire period analyses. In addition, the colour indicates the rank by the containment and health index among the analysed countries. If a country has no containment and health index record, it is coloured grey.}
\label{fig:eigen1containment}
\end{figure}

\begin{figure}[ht]
\centering

\begin{subfigure}[t]{0.48\textwidth}
    \centering
    \includegraphics[width=\linewidth,valign=m]{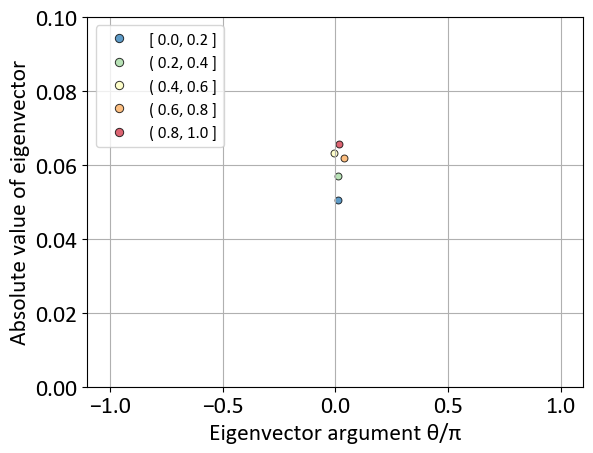}
    \caption{Entire period} 
\end{subfigure}
\begin{subfigure}[t]{0.01\textwidth}
\end{subfigure}
\begin{subfigure}[t]{0.48\textwidth}
    \centering
    \includegraphics[width=\linewidth,valign=m]{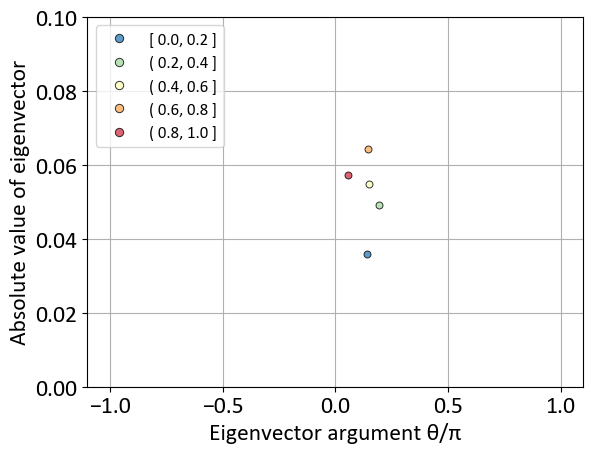}
    \caption{2020} 
\end{subfigure}
\hfill

\vspace{3ex}

\begin{subfigure}[t]{0.48\textwidth}
    \centering
    \includegraphics[width=\linewidth,valign=m]{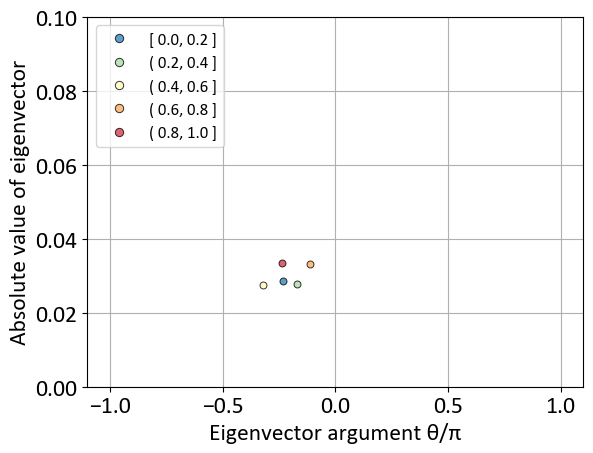}
    \caption{2021}
\end{subfigure}
\begin{subfigure}[t]{0.01\textwidth}
\end{subfigure}
\begin{subfigure}[t]{0.48\textwidth}
    \centering
    \includegraphics[width=\linewidth,valign=m]{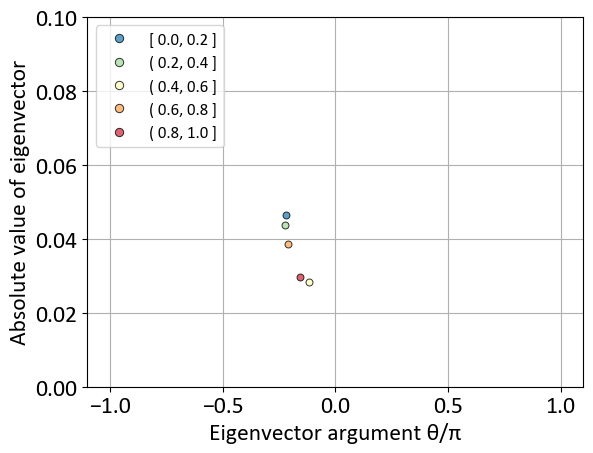}
    \caption{2022}
\end{subfigure}
\hfill

\caption{Barycentres for five groups of the containment and health index group based on rank. The abscissa corresponds to the argument, and the ordinate corresponds to the absolute value (amplitude) of the eigenvector.
Panels (a-d) represent the entire period, 2020, 2021, and 2022, respectively. Note that time progresses from right to left.
}
\label{fig:eigen1containmentbary}
\end{figure}

\begin{figure}[ht]
\centering

\begin{subfigure}[t]{0.48\textwidth}
    \centering
    \includegraphics[width=\linewidth,valign=m]{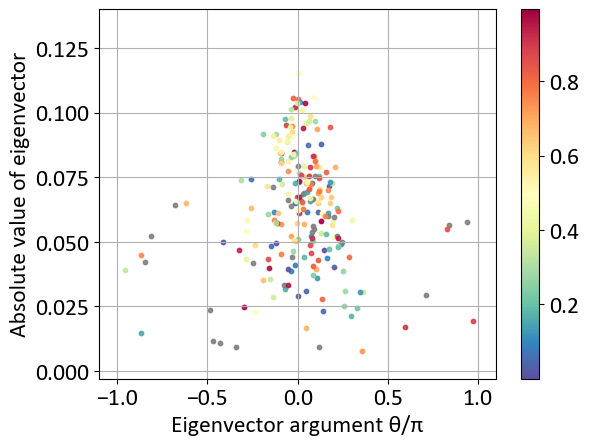}
    \caption{Entire period} 
\end{subfigure}
\begin{subfigure}[t]{0.01\textwidth}
\end{subfigure}
\begin{subfigure}[t]{0.48\textwidth}
    \centering
    \includegraphics[width=\linewidth,valign=m]{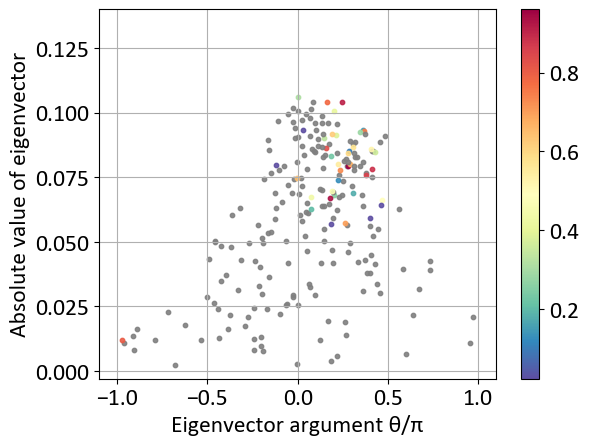}
    \caption{2020} 
\end{subfigure}
\hfill

\vspace{3ex}

\begin{subfigure}[t]{0.48\textwidth}
    \centering
    \includegraphics[width=\linewidth,valign=m]{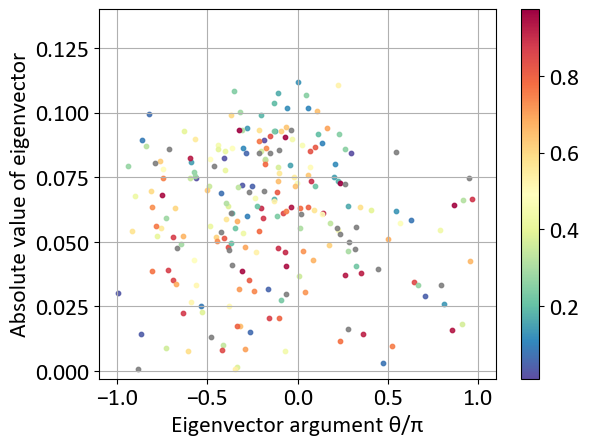}
    \caption{2021}
\end{subfigure}
\begin{subfigure}[t]{0.01\textwidth}
\end{subfigure}
\begin{subfigure}[t]{0.48\textwidth}
    \centering
    \includegraphics[width=\linewidth,valign=m]{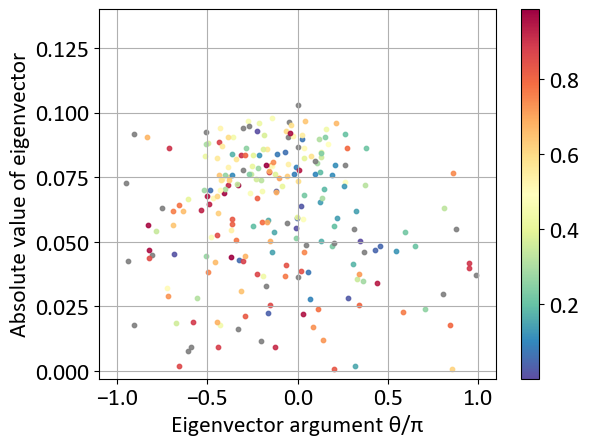}
    \caption{2022}
\end{subfigure}
\hfill

\caption{Eigenvectors, with the first eigenvalue coloured according to the vaccination rate. The abscissa corresponds to the argument, and the ordinate corresponds to the absolute value (amplitude) of the eigenvector. Note that time progresses from right to left.
Panels (a-d) represent the entire period, 2020, 2021, and 2022, respectively. Each year's vaccination rate is used for the corresponding year, while the average vaccination rate is used for the entire period. In addition, the colour indicates the rank of the countries by  vaccination rate. If a country has no vaccination record in the entire period or in 2020, it is coloured grey.}
\label{fig:eigen1vaccine}
\end{figure}

\begin{figure}[ht]
\centering

\begin{subfigure}[t]{0.48\textwidth}
    \centering
    \includegraphics[width=\linewidth,valign=m]{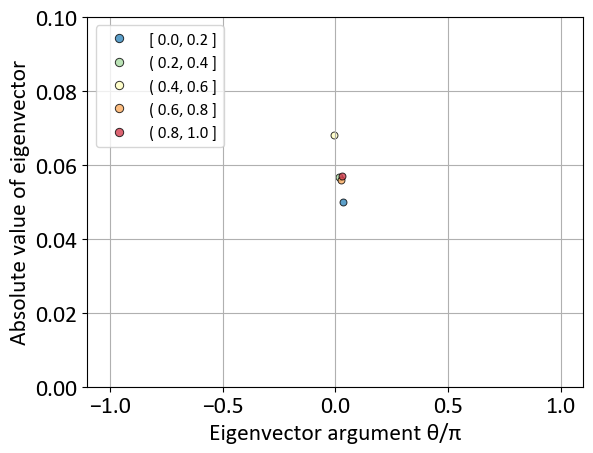}
    \caption{Entire period} 
\end{subfigure}
\begin{subfigure}[t]{0.01\textwidth}
\end{subfigure}
\begin{subfigure}[t]{0.48\textwidth}
    \centering
    \includegraphics[width=\linewidth,valign=m]{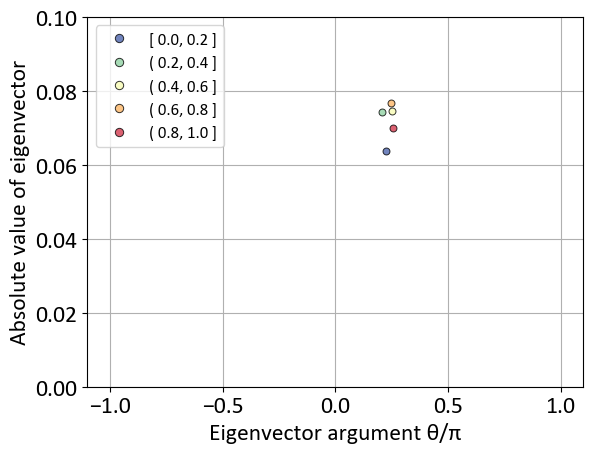}
    \caption{2020} 
\end{subfigure}
\hfill

\vspace{3ex}

\begin{subfigure}[t]{0.48\textwidth}
    \centering
    \includegraphics[width=\linewidth,valign=m]{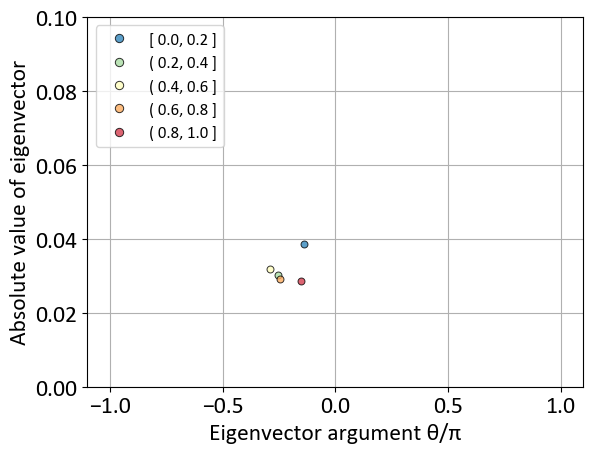}
    \caption{2021}
\end{subfigure}
\begin{subfigure}[t]{0.01\textwidth}
\end{subfigure}
\begin{subfigure}[t]{0.48\textwidth}
    \centering
    \includegraphics[width=\linewidth,valign=m]{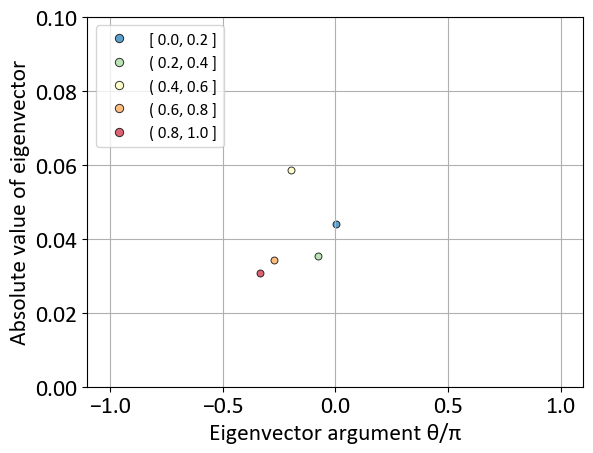}
    \caption{2022}
\end{subfigure}
\hfill

\caption{Barycentres for five groups of the vaccination rate based on rank. The abscissa corresponds to the real axis, and the ordinate corresponds to the imaginary axis.
Panels (a-d) represent the entire period, 2020, 2021, and 2022, respectively. Note that time progresses from right to left.
}
\label{fig:eigen1vacbary}
\end{figure}

\begin{figure}[ht]
\centering

\begin{subfigure}[t]{0.48\textwidth}
    \centering
    \includegraphics[width=\linewidth,valign=m]{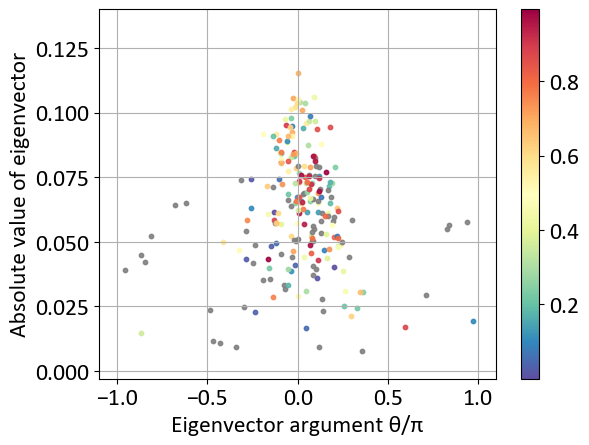}
    \caption{Entire period} 
\end{subfigure}
\begin{subfigure}[t]{0.01\textwidth}
\end{subfigure}
\begin{subfigure}[t]{0.48\textwidth}
    \centering
    \includegraphics[width=\linewidth,valign=m]{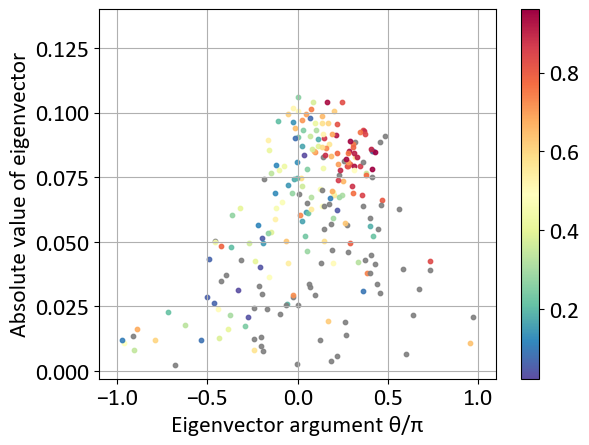}
    \caption{2020} 
\end{subfigure}
\hfill

\vspace{3ex}

\begin{subfigure}[t]{0.48\textwidth}
    \centering
    \includegraphics[width=\linewidth,valign=m]{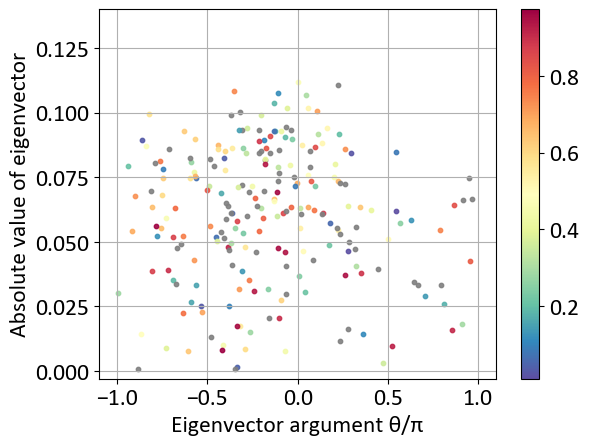}
    \caption{2021}
\end{subfigure}
\begin{subfigure}[t]{0.01\textwidth}
\end{subfigure}
\begin{subfigure}[t]{0.48\textwidth}
    \centering
    \includegraphics[width=\linewidth,valign=m]{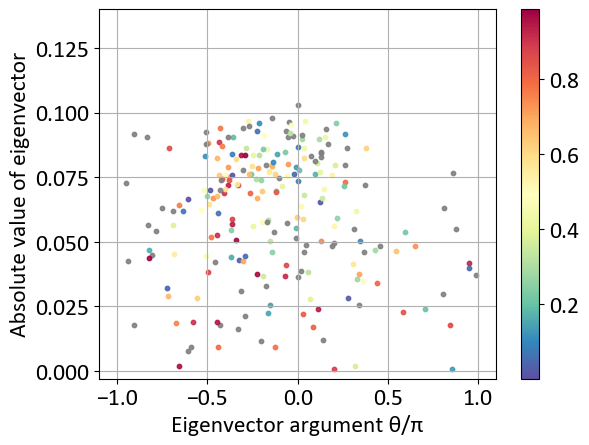}
    \caption{2022}
\end{subfigure}
\hfill

\caption{Eigenvectors, with the first eigenvalue coloured according to the democracy index. The democracy index was obtained from
the Economist Intelligence Unit \cite{economistintelligence22}.
The abscissa corresponds to the argument, and the ordinate corresponds to the absolute value (amplitude) of the eigenvector. Note that time progresses from right to left. 
Panels (a-d) represent the entire period, 2020, 2021, and 2022, respectively. Each year's democracy index data are used for the analysis of the corresponding year, while the 2020 democracy index data are used for the analysis of the entire period. In addition, the colour represents the rank of the analysed countries by the democracy index. If a country has no democracy index record, it is coloured grey.}
\label{fig:eigen1demo}
\end{figure}

\begin{figure}[ht]
\centering

\begin{subfigure}[t]{0.48\textwidth}
    \centering
    \includegraphics[width=\linewidth,valign=m]{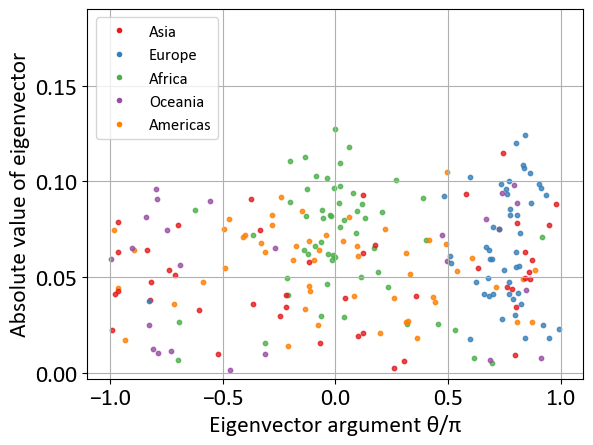}
    \caption{Region} 
\end{subfigure}
\begin{subfigure}[t]{0.01\textwidth}
\end{subfigure}
\begin{subfigure}[t]{0.48\textwidth}
    \centering
    \includegraphics[width=\linewidth,valign=m]{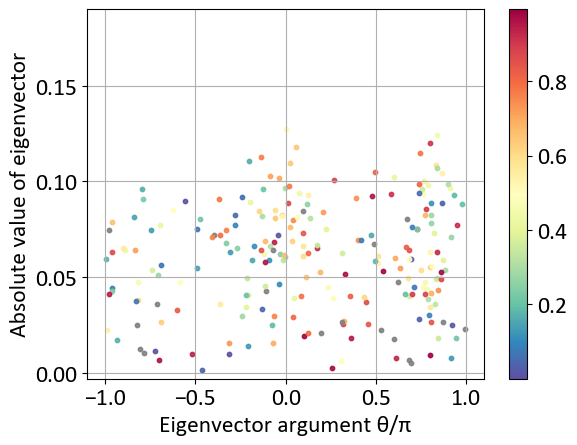}
    \caption{Population} 
\end{subfigure}
\hfill

\vspace{3ex}

\begin{subfigure}[t]{0.48\textwidth}
    \centering
    \includegraphics[width=\linewidth,valign=m]{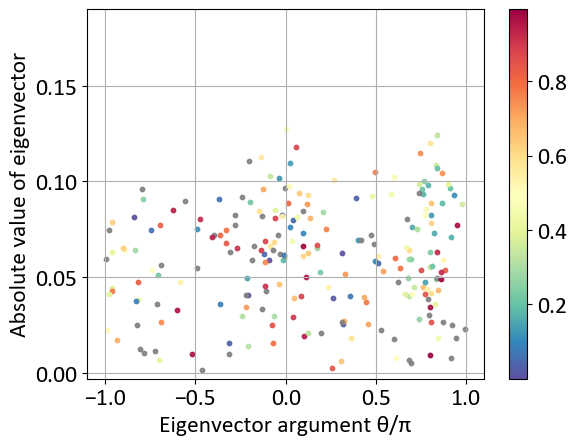}
    \caption{Stringency Index}
\end{subfigure}
\begin{subfigure}[t]{0.01\textwidth}
\end{subfigure}
\begin{subfigure}[t]{0.48\textwidth}
    \centering
    \includegraphics[width=\linewidth,valign=m]{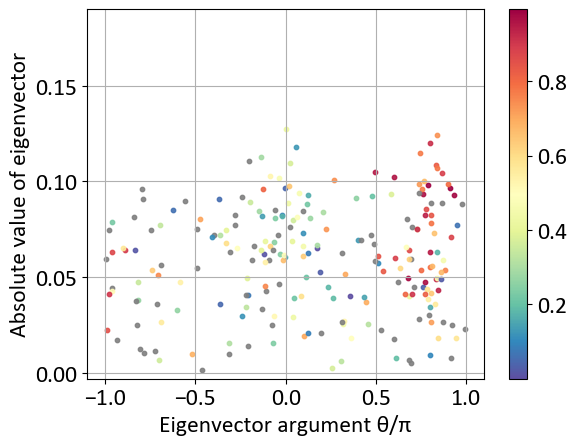}
    \caption{Democracy Index}
\end{subfigure}
\hfill

\caption{The second eigenvectors for the entire period of data. The panels are coloured by (a) region, (b) population, (c) stringency index \cite{ourworldindata23}, and (d) democracy index \cite{economistintelligence22}.
If a country has no population, stringency index, or democracy index data, it is coloured grey.
The abscissa corresponds to the argument, and the ordinate corresponds to the absolute value (amplitude) of the eigenvector. Note that time progresses from right to left. 
}
\label{fig:eigen2entire}
\end{figure}

The following is the list of figures referred to in the main text.
\begin{itemize}

\item[] SI Figure \ref{fig:numberofinfected}: Number of infected cases in all countries. In the main text, only the 20 most populous countries are shown for improved visibility.

\item[] SI Figure \ref{fig:logratio}: Log ratio of infected cases in all countries. In the main text, only the 20 most populous countries are shown for improved visibility.

\item[] SI Figure \ref{fig:legend}: The legends for SI Figures \ref{fig:numberofinfected} and \ref{fig:logratio}.

\item[] SI Figure \ref{fig:aux}: Pair plots between auxiliary data points.

\item[] SI Figure \ref{fig:scree}: Scree plots for each year. In the main text, the scree plot for the entire period is shown.

\item[] SI Figure \ref{fig:eigen1region}: The first eigenvectors for the entire period, 2020, 2021, and 2022. The plots are coloured by region. The axes represent the argument and the ordinate.

\item[] SI Figure \ref{fig:eigen1regioncomplex}: The first eigenvectors for the entire period, 2020, 2021, and 2022. The plots are coloured by region. The figure shown on the complex plane corresponds to SI Figure \ref{fig:eigen1region} in the main text.

\item[] SI Figure \ref{fig:eigen1population}: The first eigenvectors for the entire period, 2020, 2021, and 2022. The plots are coloured according to population. The axes represent the argument and the ordinate.

\item[] SI Figure \ref{fig:eigen1gdpcapita}: The first eigenvectors for the entire period, 2020, 2021, and 2022. The plots are coloured according to GDP per capita. The axes represent the argument and the ordinate.

\item[] SI Figure \ref{fig:eigen1gdpcapitabary}: Barycentres for each of five GDP per capita group based on rank. The panels show the entire period, 2020, 2021, and 2022. This figure corresponds to SI Figure \ref{fig:eigen1gdpcapita}.

\item[] SI Figure \ref{fig:eigen1stringency}: The first eigenvectors for the entire period, 2020, 2021, and 2022. The plots are coloured according to the stringency index \cite{ourworldindata23}.
The axes represent the argument and the ordinate.

\item[] SI Figure \ref{fig:eigen1stringencybary}: Barycentres for each of five stringency index group based on rank. The panels show the entire period, 2020, 2021, and 2022. This figure corresponds to SI Figure \ref{fig:eigen1stringency}.

\item[] SI Figure \ref{fig:eigen1containment}: The first eigenvectors for the entire period, 2020, 2021, and 2022. The plots are coloured according to the containment and health index \cite{ourworldindata23}.
The axes represent the argument and the ordinate.

\item[] SI Figure \ref{fig:eigen1containmentbary}: Barycentres for each of five containment and health index groups based on rank. The panels show the entire period, 2020, 2021, and 2022. This figure corresponds to SI Figure \ref{fig:eigen1containment}.

\item[] SI Figure \ref{fig:eigen1vaccine}: The first eigenvectors for the entire period, 2020, 2021, and 2022. The plots are coloured according to the vaccination rate. The axes represent the argument and the ordinate.

\item[] SI Figure \ref{fig:eigen1vacbary}: Barycentres for each of five vaccination rate groups based on rank. The panels show the entire period, 2020, 2021, and 2022. This figure corresponds to SI Figure \ref{fig:eigen1vaccine}.

\item[] SI Figure \ref{fig:eigen1demo}: The first eigenvectors for the entire period, 2020, 2021, and 2022. The plots are coloured according to the democracy index. The figure shown on the complex plane corresponds to Figure 5 in the main text.

\item[] SI Figure \ref{fig:eigen2entire}: The second eigenvectors for the entire period data. The panels are coloured by region, population, stringency index, and democracy index.

\item[] SI Figure \ref{fig:eigen22020}: The second eigenvectors for the 2020 data. The panels are coloured by region, population, stringency index, and democracy index.

\item[] SI Figure \ref{fig:eigen22021}: The second eigenvectors for the 2021 data. The panels are coloured by region, population, stringency index, and democracy index.

\item[] SI Figure \ref{fig:eigen22022}: The second eigenvectors for the 2022 data. The panels are coloured by region, population, stringency index, and democracy index.

\end{itemize}


\clearpage

\clearpage

\clearpage

\clearpage

\clearpage

\clearpage

\clearpage

\clearpage

\clearpage

\clearpage

\clearpage

\clearpage

\clearpage

\begin{figure}[ht]
\centering

\begin{subfigure}[t]{0.48\textwidth}
    \centering
    \includegraphics[width=\linewidth,valign=m]{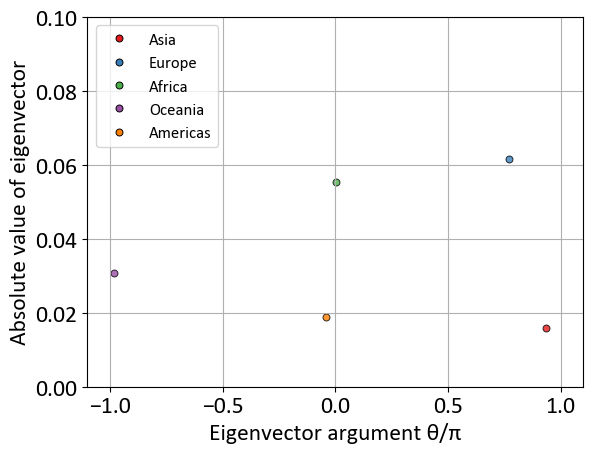}
    \caption{Region} 
\end{subfigure}
\begin{subfigure}[t]{0.01\textwidth}
\end{subfigure}
\begin{subfigure}[t]{0.48\textwidth}
    \centering
    \includegraphics[width=\linewidth,valign=m]{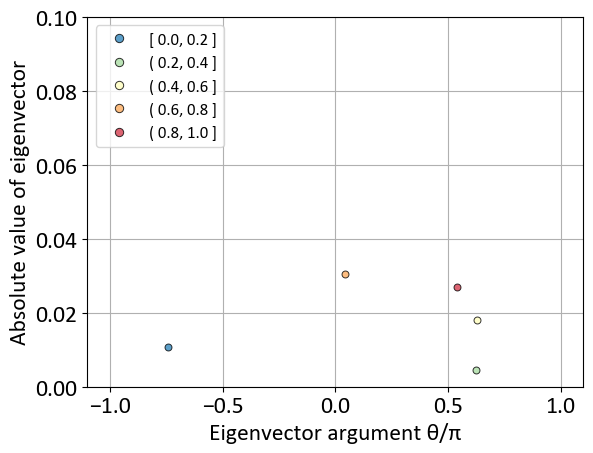}
    \caption{Population} 
\end{subfigure}
\hfill

\vspace{3ex}

\begin{subfigure}[t]{0.48\textwidth}
    \centering
    \includegraphics[width=\linewidth,valign=m]{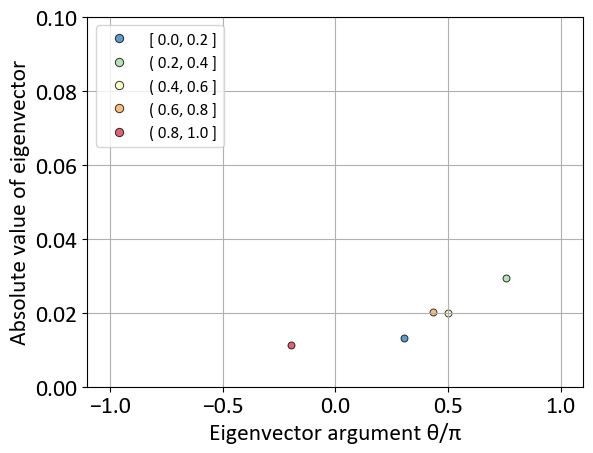}
    \caption{Stringency Index}
\end{subfigure}
\begin{subfigure}[t]{0.01\textwidth}
\end{subfigure}
\begin{subfigure}[t]{0.48\textwidth}
    \centering
    \includegraphics[width=\linewidth,valign=m]{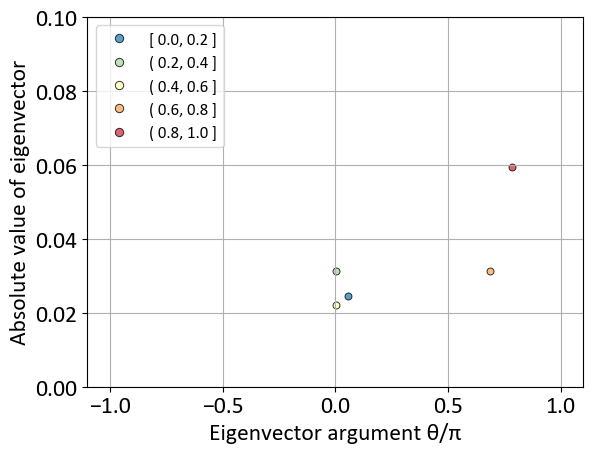}
    \caption{Democracy Index}
\end{subfigure}
\hfill

\caption{
Barycentres of the second eigenvectors for data of the entire period. The panels are coloured by (a) region, (b) population, (c) stringency index \cite{ourworldindata23}, and (d) democracy index \cite{economistintelligence22}.
The abscissa corresponds to the argument, and the ordinate corresponds to the absolute value (amplitude) of the eigenvector. Note that time progresses from right to left. 
}
\label{fig:eigen2entirebary}
\end{figure}

\clearpage

\begin{figure}[ht]
\centering

\begin{subfigure}[t]{0.48\textwidth}
    \centering
    \includegraphics[width=\linewidth,valign=m]{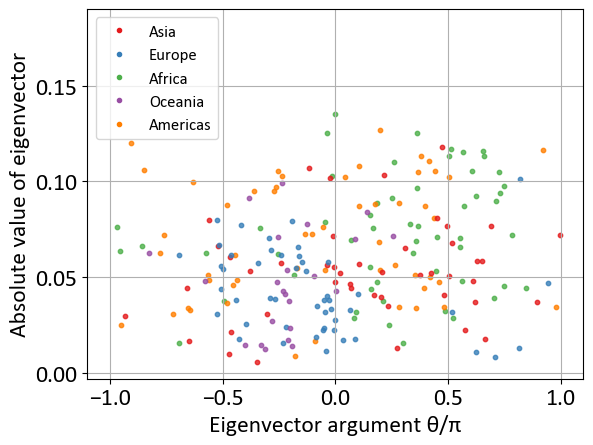}
    \caption{Region} 
\end{subfigure}
\begin{subfigure}[t]{0.01\textwidth}
\end{subfigure}
\begin{subfigure}[t]{0.48\textwidth}
    \centering
    \includegraphics[width=\linewidth,valign=m]{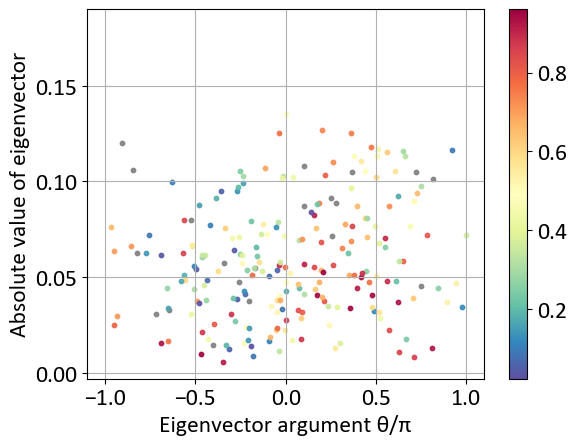}
    \caption{Population} 
\end{subfigure}
\hfill

\vspace{3ex}

\begin{subfigure}[t]{0.48\textwidth}
    \centering
    \includegraphics[width=\linewidth,valign=m]{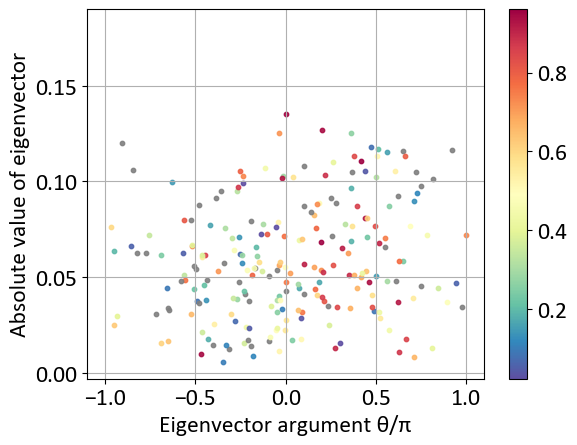}
    \caption{Stringency Index}
\end{subfigure}
\begin{subfigure}[t]{0.01\textwidth}
\end{subfigure}
\begin{subfigure}[t]{0.48\textwidth}
    \centering
    \includegraphics[width=\linewidth,valign=m]{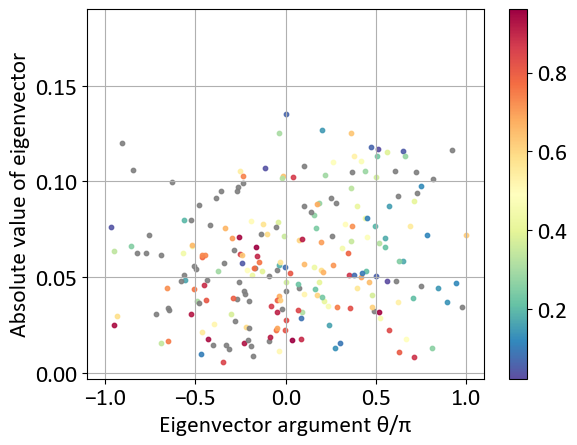}
    \caption{Democracy Index}
\end{subfigure}
\hfill

\caption{The second eigenvector for the 2020 data. The panels are coloured by (a) region, (b) population, (c) stringency index \cite{ourworldindata23}, and (d) democracy index \cite{economistintelligence22}.
If a country has no population, stringency index, or democracy index data, it is coloured grey.
The abscissa corresponds to the real axis, and the ordinate corresponds to the imaginary axis.
Note that time progresses from right to left.
}
\label{fig:eigen22020}
\end{figure}

\clearpage

\begin{figure}[ht]
\centering

\begin{subfigure}[t]{0.48\textwidth}
    \centering
    \includegraphics[width=\linewidth,valign=m]{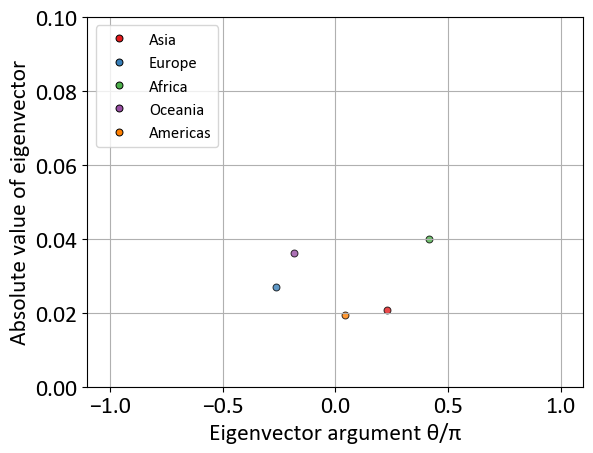}
    \caption{Region} 
\end{subfigure}
\begin{subfigure}[t]{0.01\textwidth}
\end{subfigure}
\begin{subfigure}[t]{0.48\textwidth}
    \centering
    \includegraphics[width=\linewidth,valign=m]{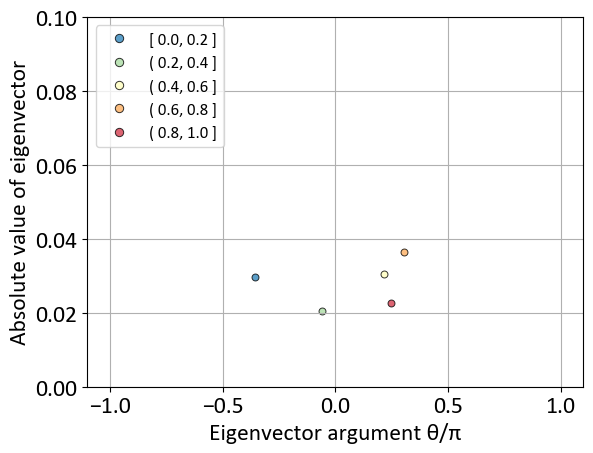}
    \caption{Population} 
\end{subfigure}
\hfill

\vspace{3ex}

\begin{subfigure}[t]{0.48\textwidth}
    \centering
    \includegraphics[width=\linewidth,valign=m]{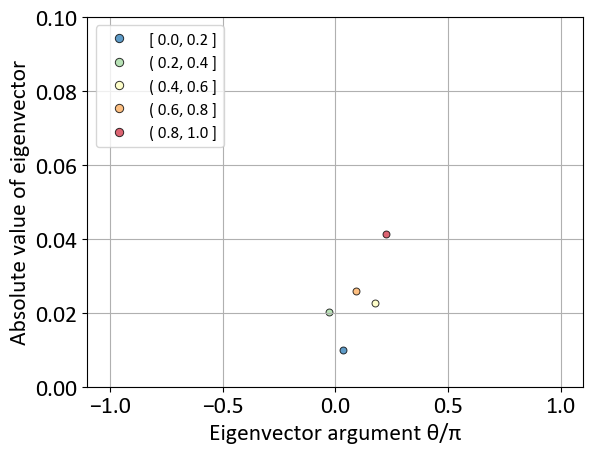}
    \caption{Stringency Index}
\end{subfigure}
\begin{subfigure}[t]{0.01\textwidth}
\end{subfigure}
\begin{subfigure}[t]{0.48\textwidth}
    \centering
    \includegraphics[width=\linewidth,valign=m]{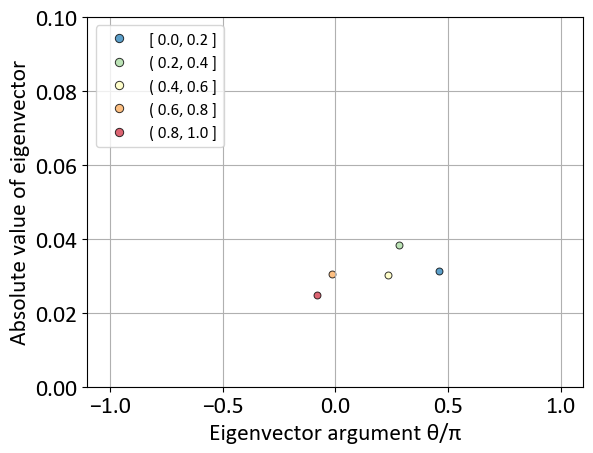}
    \caption{Democracy Index}
\end{subfigure}
\hfill

\caption{
Barycentres of the second eigenvectors for the 2020 data. The panels are coloured by (a) region, (b) population, (c) stringency index \cite{ourworldindata23}, and (d) democracy index \cite{economistintelligence22}.
The abscissa corresponds to the argument, and the ordinate corresponds to the absolute value (amplitude) of the eigenvector. Note that time progresses from right to left. 
}
\label{fig:eigen22020bary}
\end{figure}

\clearpage

\begin{figure}[ht]
\centering

\begin{subfigure}[t]{0.48\textwidth}
    \centering
    \includegraphics[width=\linewidth,valign=m]{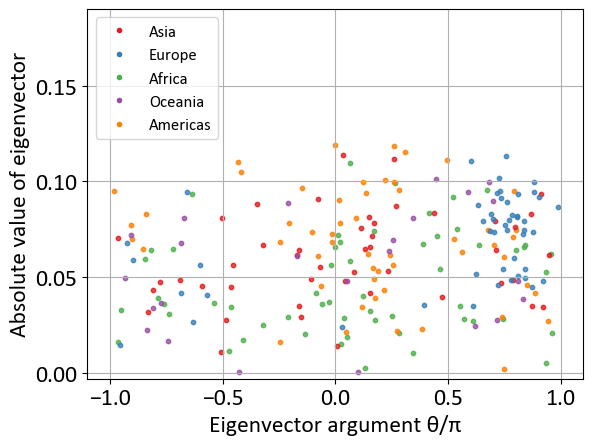}
    \caption{Region} 
\end{subfigure}
\begin{subfigure}[t]{0.01\textwidth}
\end{subfigure}
\begin{subfigure}[t]{0.48\textwidth}
    \centering
    \includegraphics[width=\linewidth,valign=m]{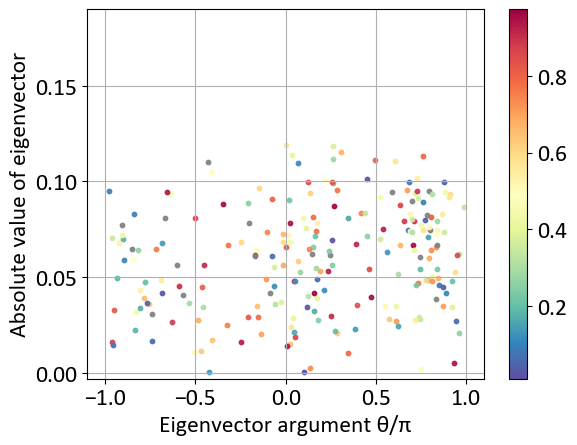}
    \caption{Population} 
\end{subfigure}
\hfill

\vspace{3ex}

\begin{subfigure}[t]{0.48\textwidth}
    \centering
    \includegraphics[width=\linewidth,valign=m]{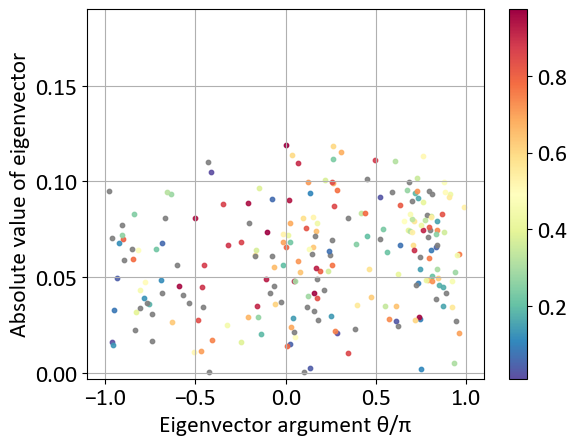}
    \caption{Stringency Index}
\end{subfigure}
\begin{subfigure}[t]{0.01\textwidth}
\end{subfigure}
\begin{subfigure}[t]{0.48\textwidth}
    \centering
    \includegraphics[width=\linewidth,valign=m]{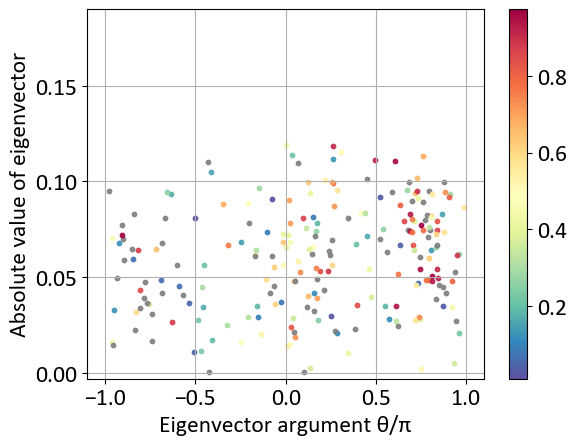}
    \caption{Democracy Index}
\end{subfigure}
\hfill

\caption{The second eigenvectors for the 2021 data. The panels are coloured by region, population, stringency index \cite{ourworldindata23}, and democracy index \cite{economistintelligence22}.
If a country has no population, stringency index, or democracy index data, it is coloured grey.
The abscissa corresponds to the real axis, and the ordinate corresponds to the imaginary axis.
Note that time progresses from right to left.
}
\label{fig:eigen22021}
\end{figure}

\clearpage

\begin{figure}[ht]
\centering

\begin{subfigure}[t]{0.48\textwidth}
    \centering
    \includegraphics[width=\linewidth,valign=m]{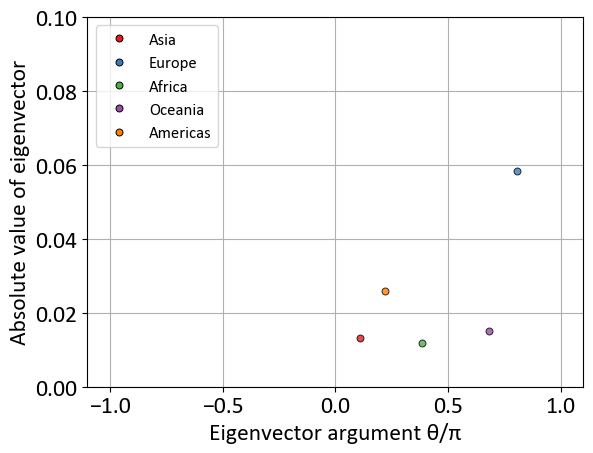}
    \caption{Region} 
\end{subfigure}
\begin{subfigure}[t]{0.01\textwidth}
\end{subfigure}
\begin{subfigure}[t]{0.48\textwidth}
    \centering
    \includegraphics[width=\linewidth,valign=m]{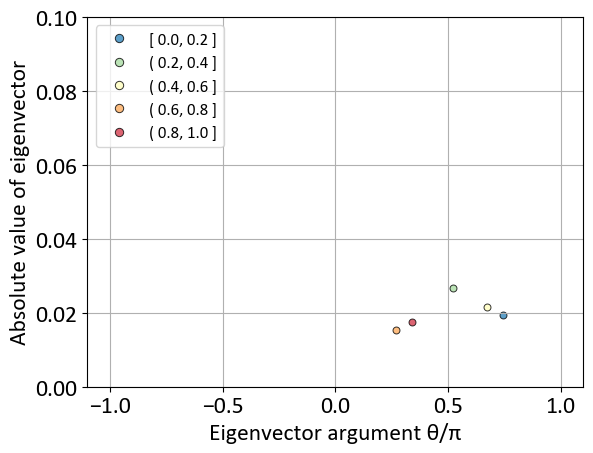}
    \caption{Population} 
\end{subfigure}
\hfill

\vspace{3ex}

\begin{subfigure}[t]{0.48\textwidth}
    \centering
    \includegraphics[width=\linewidth,valign=m]{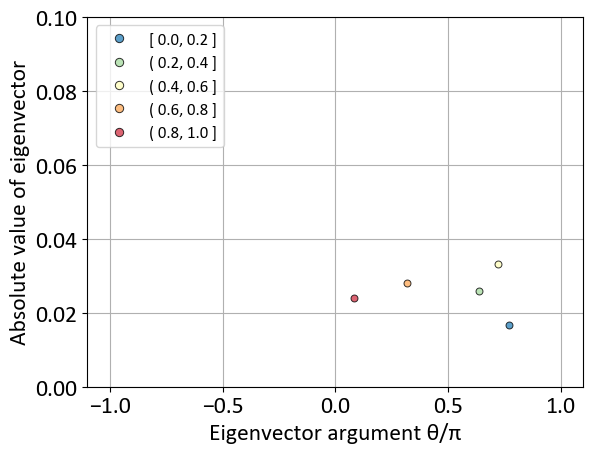}
    \caption{Stringency Index}
\end{subfigure}
\begin{subfigure}[t]{0.01\textwidth}
\end{subfigure}
\begin{subfigure}[t]{0.48\textwidth}
    \centering
    \includegraphics[width=\linewidth,valign=m]{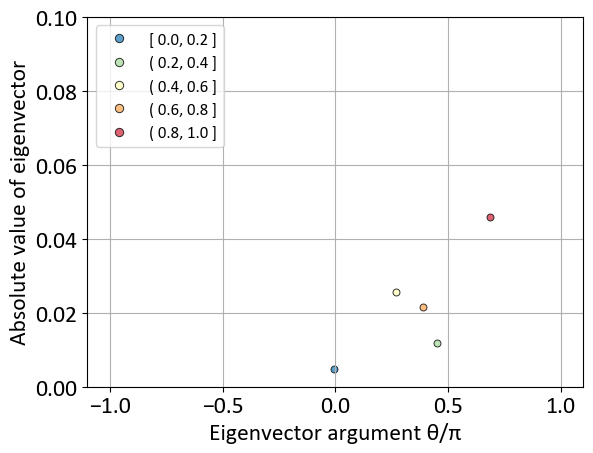}
    \caption{Democracy Index}
\end{subfigure}
\hfill

\caption{
Barycentres of the second eigenvectors for the 2021 data. The panels are coloured by (a) region, (b) population, (c) stringency index \cite{ourworldindata23}, and (d) democracy index \cite{economistintelligence22}.
The abscissa corresponds to the argument, and the ordinate corresponds to the absolute value (amplitude) of the eigenvector. Note that time progresses from right to left. 
}
\label{fig:eigen22021bary}
\end{figure}

\clearpage

\begin{figure}[ht]
\centering

\begin{subfigure}[t]{0.48\textwidth}
    \centering
    \includegraphics[width=\linewidth,valign=m]{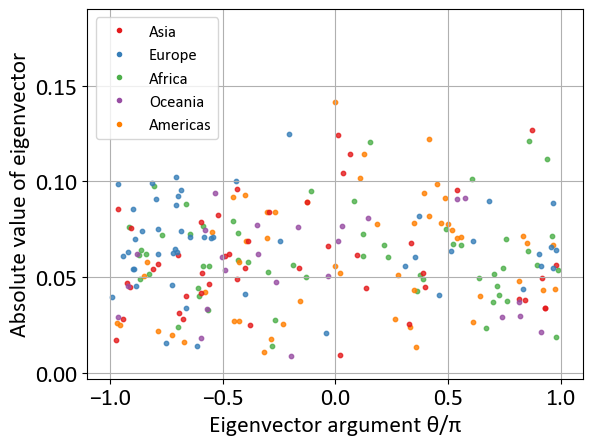}
    \caption{Region} 
\end{subfigure}
\begin{subfigure}[t]{0.01\textwidth}
\end{subfigure}
\begin{subfigure}[t]{0.48\textwidth}
    \centering
    \includegraphics[width=\linewidth,valign=m]{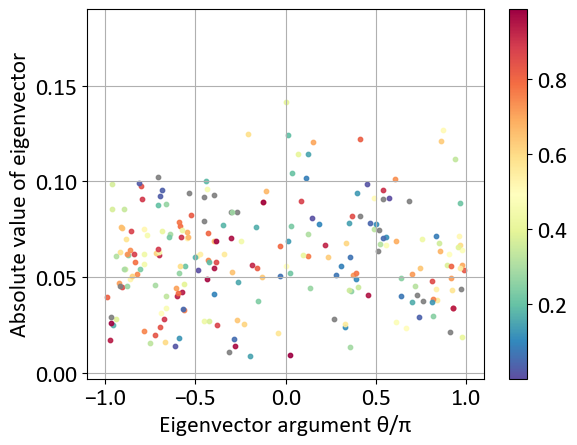}
    \caption{Population} 
\end{subfigure}
\hfill

\vspace{3ex}

\begin{subfigure}[t]{0.48\textwidth}
    \centering
    \includegraphics[width=\linewidth,valign=m]{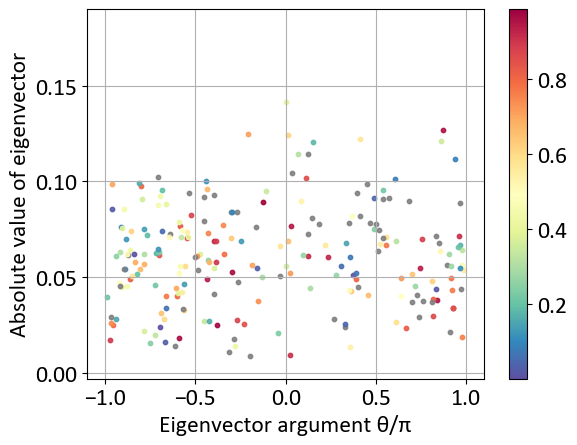}
    \caption{Stringency Index}
\end{subfigure}
\begin{subfigure}[t]{0.01\textwidth}
\end{subfigure}
\begin{subfigure}[t]{0.48\textwidth}
    \centering
    \includegraphics[width=\linewidth,valign=m]{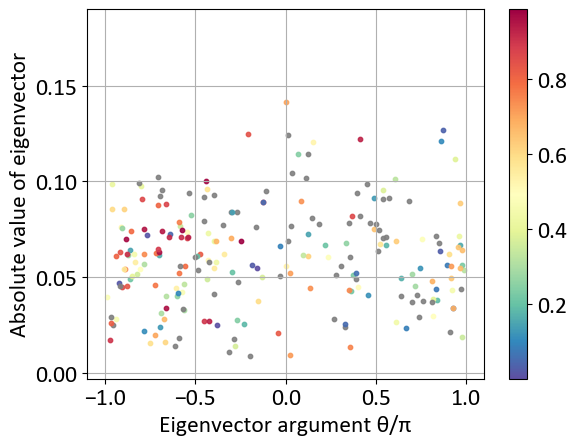}
    \caption{Democracy Index}
\end{subfigure}
\hfill

\caption{The second eigenvectors for the 2022 data. The panels are coloured by region, population, stringency index \cite{ourworldindata23}, and democracy index \cite{economistintelligence22}.
If a country has no population, stringency index, or democracy index data, it is coloured grey.
The abscissa corresponds to the real axis, and the ordinate corresponds to the imaginary axis.
Note that time progresses from right to left.
}
\label{fig:eigen22022}
\end{figure}

\clearpage

\begin{figure}[ht]
\centering

\begin{subfigure}[t]{0.48\textwidth}
    \centering
    \includegraphics[width=\linewidth,valign=m]{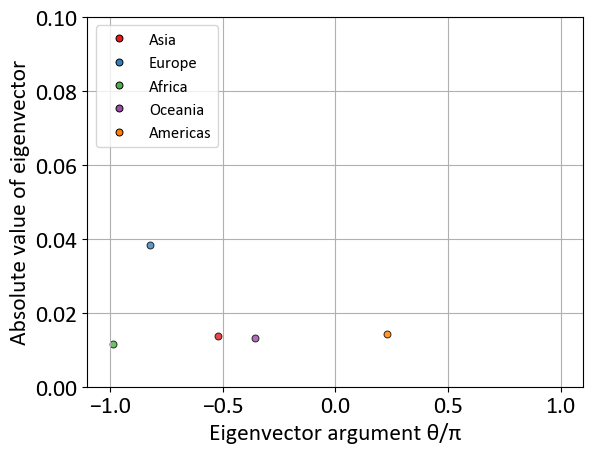}
    \caption{Region} 
\end{subfigure}
\begin{subfigure}[t]{0.01\textwidth}
\end{subfigure}
\begin{subfigure}[t]{0.48\textwidth}
    \centering
    \includegraphics[width=\linewidth,valign=m]{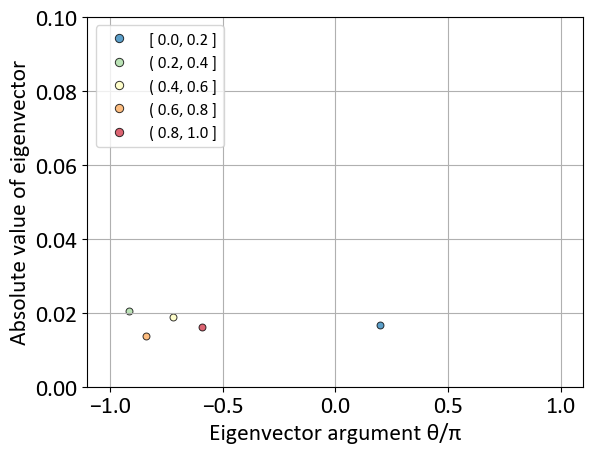}
    \caption{Population} 
\end{subfigure}
\hfill

\vspace{3ex}

\begin{subfigure}[t]{0.48\textwidth}
    \centering
    \includegraphics[width=\linewidth,valign=m]{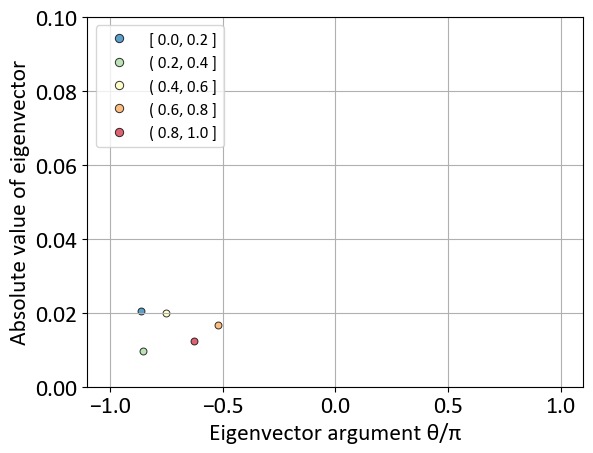}
    \caption{Stringency Index}
\end{subfigure}
\begin{subfigure}[t]{0.01\textwidth}
\end{subfigure}
\begin{subfigure}[t]{0.48\textwidth}
    \centering
    \includegraphics[width=\linewidth,valign=m]{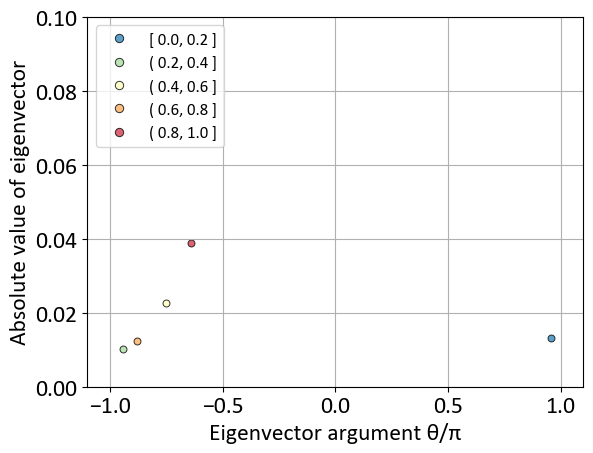}
    \caption{Democracy Index}
\end{subfigure}
\hfill

\caption{
Barycentres of the second eigenvectors for the 2022 data. The panels are coloured by (a) region, (b) population, (c) stringency index \cite{ourworldindata23}, and (d) democracy index \cite{economistintelligence22}.
The abscissa corresponds to the argument, and the ordinate corresponds to the absolute value (amplitude) of the eigenvector. Note that time progresses from right to left. 
}
\label{fig:eigen22022bary}
\end{figure}

\clearpage

\begin{figure}[ht]
\centering

\begin{subfigure}[t]{0.48\textwidth}
    \centering
    \includegraphics[width=\linewidth,valign=m]{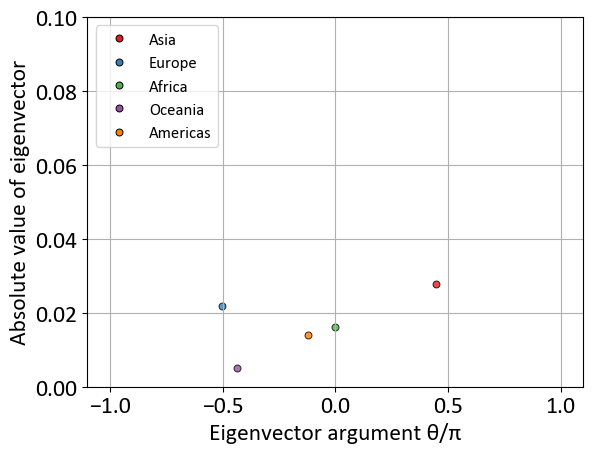}
    \caption{Region} 
\end{subfigure}
\begin{subfigure}[t]{0.01\textwidth}
\end{subfigure}
\begin{subfigure}[t]{0.48\textwidth}
    \centering
    \includegraphics[width=\linewidth,valign=m]{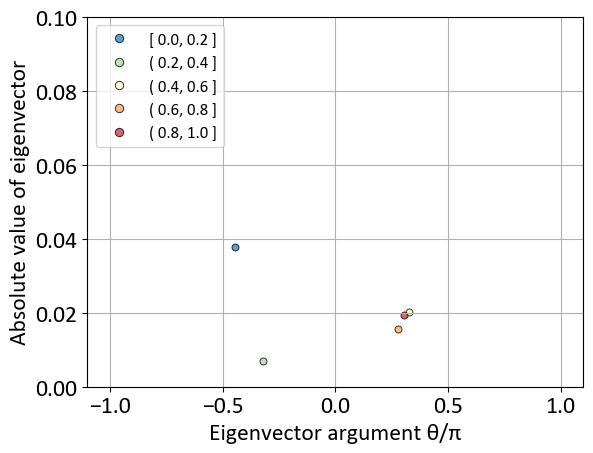}
    \caption{Population} 
\end{subfigure}
\hfill

\vspace{3ex}

\begin{subfigure}[t]{0.48\textwidth}
    \centering
    \includegraphics[width=\linewidth,valign=m]{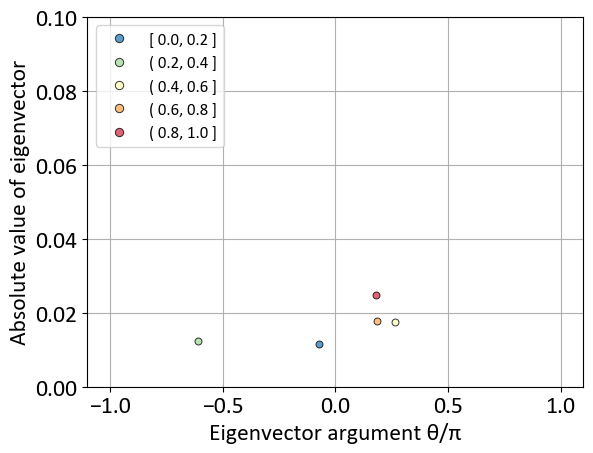}
    \caption{Stringency Index}
\end{subfigure}
\begin{subfigure}[t]{0.01\textwidth}
\end{subfigure}
\begin{subfigure}[t]{0.48\textwidth}
    \centering
    \includegraphics[width=\linewidth,valign=m]{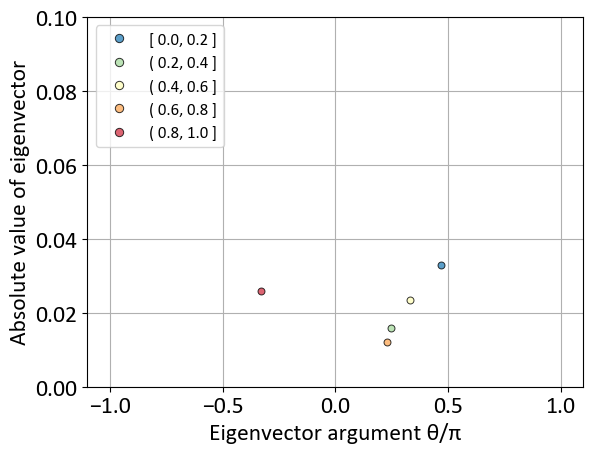}
    \caption{Democracy Index}
\end{subfigure}
\hfill

\caption{
Barycentres of the third eigenvectors for the data of the entire period. The panels are coloured by (a) region, (b) population, (c) stringency index \cite{ourworldindata23}, and (d) democracy index \cite{economistintelligence22}.
The abscissa corresponds to the argument, and the ordinate corresponds to the absolute value (amplitude) of the eigenvector. Note that time progresses from right to left. 
}
\label{fig:eigen3entirebary}
\end{figure}

\clearpage

\section{Supplementary tables}

The following list contains the tables referred to in the main text.

\begin{itemize}
\item[] SI Table \ref{tbl:sample}: Sample size of the auxiliary data.
\item[] SI Table \ref{tbl:barycenter}: Mean distance from the barycentre. The mean distance is calculated for each group.
\end{itemize}

\begin{table}[ht]

\caption{Sample sizes of auxiliary data. The numbers indicates the number of observed countries in each dataset.}
\label{tbl:sample}
\center

\begin{tabular}{lccc}
\hline
  & 2020 & 2021 & 2022 \\
\hline Population & 265 & 265 &  265 \\
 GDP & 257 & 251 & 233 \\
 Stringency Index & 183 & 180 & 180 \\
 Containment Index & 184 & 180 & 180 \\
 Vaccinations Rate & 39 & 217 & 213 \\
 Democracy Index & 167 & 167 & 167 \\
\hline
\end{tabular}
\end{table}

\begin{table}[ht]
\center

\caption{Mean distance from the barycentre. The mean distance is calculated for each group. The columns indicate the position within the five groups, which are divided by rank.}
\label{tbl:barycenter}

\begin{tabular}{ll|llll}
 & Group & Entire & 2020  & 2021  & 2022  \\
 &  & Period & & &  \\
 \hline
Regions & Asia & 0.030  & 0.044  & 0.046  & 0.041  \\
 & Europe & 0.027  & 0.030  & 0.053  & 0.038  \\
 & Africa & 0.027  & 0.040  & 0.055  & 0.044  \\
 & Oceania & 0.046  & 0.031  & 0.057  & 0.053  \\
 & Americas & 0.026  & 0.041  & 0.053  & 0.055  \\
 \hline
Population &  First  & 0.036  & 0.034  & 0.054  & 0.058  \\
 &  Second & 0.032  & 0.042  & 0.053  & 0.049  \\
 &  Third & 0.029  & 0.043  & 0.054  & 0.047  \\
 &  Fourth & 0.028  & 0.044  & 0.055  & 0.047  \\
 &  Fifth  & 0.032  & 0.044  & 0.053  & 0.050  \\
 \hline
GDP/Population &  First  & 0.025  & 0.035  & 0.056  & 0.040  \\
 &  Second & 0.041  & 0.043  & 0.054  & 0.058  \\
 &  Third & 0.032  & 0.042  & 0.048  & 0.046  \\
 &  Fourth & 0.027  & 0.037  & 0.058  & 0.048  \\
 &  Fifth  & 0.022  & 0.030  & 0.046  & 0.043  \\
 \hline
Stringency Index &  First  & 0.028  & 0.046  & 0.063  & 0.038  \\
 &  Second & 0.032  & 0.038  & 0.050  & 0.046  \\
 &  Third & 0.032  & 0.048  & 0.053  & 0.048  \\
 &  Fourth & 0.031  & 0.044  & 0.054  & 0.055  \\
 &  Fifth  & 0.030  & 0.039  & 0.047  & 0.059  \\
 \hline
Containment &  First  & 0.033  & 0.044  & 0.062  & 0.044  \\
\& Health Index &  Second & 0.030  & 0.045  & 0.057  & 0.044  \\
 &  Third & 0.031  & 0.044  & 0.043  & 0.051  \\
 &  Fourth & 0.031  & 0.044  & 0.055  & 0.047  \\
 &  Fifth  & 0.030  & 0.041  & 0.050  & 0.061  \\
 \hline
Vaccination Rate &  First  & 0.028  & 0.038  & 0.062  & 0.045  \\
 &  Second & 0.034  & 0.032  & 0.056  & 0.050  \\
 &  Third & 0.029  & 0.028  & 0.056  & 0.051  \\
 &  Fourth & 0.031  & 0.022  & 0.050  & 0.054  \\
 &  Fifth  & 0.032  & 0.035  & 0.045  & 0.047  \\
 \hline
Democracy Index &  First  & 0.032  & 0.041  & 0.059  & 0.050  \\
 &  Second & 0.029  & 0.040  & 0.049  & 0.047  \\
 &  Third & 0.035  & 0.041  & 0.059  & 0.048  \\
 &  Fourth & 0.026  & 0.038  & 0.055  & 0.049  \\
 &  Fifth  & 0.023  & 0.025  & 0.045  & 0.040  \\
\hline
\end{tabular}

\end{table}

\clearpage

\section{Analyses of the second eigenvectors}

The second eigenvector for the 2020 data shows wide ranges for lead and lag, in contrast to the first eigenvector for the 2020 data (SI Figure 21). In terms of regions, the results are entirely different from those of the first eigenvector, with Africa leading and Europe lagging. This suggests that even during this period, Africa and Europe already show leading and lagging trends, respectively, as observed from 2021 to 2022. In terms of population, a lead-lag relationship can be observed, with countries with larger populations leading; however, population does not significantly affect the amplitude. The same results are found for the stringency index. Regarding the democracy index, the lead-lag relationships and amplitudes are completely opposite those observed with the first eigenvector. The leading trend is associated with a lower democracy index, and the lagging trend is associated with a higher democracy index. Moreover, a larger amplitude is associated with a lower democracy index, and a smaller amplitude is associated with a higher democracy index.

The second eigenvector for the 2021 data shows that the spread in the argument is the same as that for the first eigenvector (SI Figure 22), but there are some differences in the results. 
The first eigenvector does not show outstanding features in region or democracy index, but the second eigenvector captures the leading trend of European countries and a high democracy index. There are no distinctive results for the population or stringency indeces, similar to the results of the analyses for the first eigenvector.